\documentclass[12pt]{article}

\usepackage{graphics}
\usepackage{epsfig}
\usepackage{times}
\usepackage{amsmath}
\usepackage{url}
\usepackage{rotating}
\usepackage{subfigure}

\title{{\bf MyZone: A Next-Generation Online Social Network} \\
\it Tech. Report}
\author{ {\bf Alireza Mahdian, John Black, Richard Han and Shivakant Mishra}  \\
Department of Computer Science \\
University of Colorado at Boulder\\
{\small alireza.mahdian@colorado.edu}
}
\date{}

\begin{document}
\pagestyle{plain}
\pagenumbering{roman}
\maketitle

\pagebreak
\begin{abstract}

This technical report considers the design of a social network that would address the shortcomings of the current ones, and identifies user privacy, security, and  service availability as strong motivations that push the architecture of the proposed design to be distributed.   We describe our design in detail and identify the property of \emph{resiliency} as a key objective for the overall design philosophy. 

We define the system goals, threat model, and trust model as part of the system model, and discuss the challenges in adapting such distributed frameworks to become highly available and highly resilient in potentially hostile environments.  We propose a distributed solution to address these challenges based on a trust-based friendship model for replicating user profiles and disseminating messages, and examine how this approach builds upon prior work in distributed Peer-to-Peer (P2P) networks.

\end{abstract}

\pagebreak
\tableofcontents
\pagebreak

\cleardoublepage
\pagenumbering{arabic}

\section{Introduction}
\label{ch:intro}

Social network is a form of social structure that consists of users connected to each other based on a set of commonalities. An Online social network (OSN) facilitates a social network using online services. The most notable feature of OSNs is the possibility of remote interactions among users. In addition, most of them provide a variety of other features, from facilitating users to expand their social circles through common interests, mutual friends, or even searching for long lost acquaintances, to sharing different contents with different users either individually or as groups, to allowing users to reflect interest or opinion on shared contents, and much more.

One advantage of OSNs is their high reachability which is a byproduct of human behavior: people tend to be medium for each other. Therefore, OSNs would provide users with an even broader audience than just their immediate friends. This has enabled OSNs to emerge as a pervasive and almost ubiquitous social media. As a result, their social impact has been demonstrated around the globe both at the local and global scales. These features have had a huge impact on the overwhelming popularity of OSNs. Due to these appealing features OSNs have enjoyed huge success in attracting loyal users.

The most successful OSNs have been free of charge for their users. The assets of an OSN service provider comes from its number of users, which in turn translates to its influence. The profit earned by an OSN service provider comes from the value of the user information. OSN service providers have complete access to user data, in some cases, even long after they leave the OSN. This enables them to profile their users individually, and target them with more specific ads, based on the history of their interactions. This has essentially made OSNs a very successful media for effective advertising. In this respect, OSN users are both its consumers and products, i.e. they use its seemingly free services but on the other hand they provide information that can be used for effective targeted advertisement. In addition, user information can also be sold to third parties. 

Although, users of most OSN services have to agree to a set of privacy policies, dictated by the OSN service provider, most people blindly agree to them and implicitly allow the OSN service provider to abuse their information by selling them to market researchers and advertisers, which will ultimately be used for targeted advertisment. In more extreme cases, some third party applications have been responsible for causing security breaches by installing adware and spyware. To make matters worse, continuous changes in the privacy policies of OSNs have resulted in frustrated users who (perhaps rightly so) are becoming more suspicious. 

A recent survey has indicated that more and more people are taking precautions when it comes to using OSN services\cite{survey}.  This user privacy violation is the most serious shortcoming of conventional OSNs and stems from a centralized server architecture that enables a single entity i.e. the OSN service provider, to have complete access to all of its users. Later in this technical report, we introduce an alternative distributed architecture, in which users own their data, and have it hosted only by a number of trusted peers, i.e. friends.

A second shortcoming that many OSN users are facing today is, that of government imposed censorship. OSNs and social media are in widespread use today as a political tool, giving people a voice and a way to express, and spread their opinions freely and effectively. In some cases, especially in those parts of the world that have yet to enjoy free media, this has led to popular uprisings. In light of recent events in the Middle East and North Africa, social networks such as Twitter\cite{twitter} and Facebook\cite{facebook} have been (perhaps rightly) credited as one of the main contributors to political changes in Egypt and Tunisia\cite{nytimes, cnn}. This has caused many concerned governments to censor OSN services\cite{cnn_egypt_censor, washingtonpost, nytimes_china_censor}. This can be easily achieved either by traffic filtering or implementing attacks on the OSN service provider or even asking the providers to block their services to an entire nation\cite{cnn_government}.

The centralized server architecture of current OSNs is thus quite brittle and vulnerable to denial-of-service attacks that simply pull the connection to the server.  Some measure of distribution is possible, i.e. Facebook can have many points of presence spread throughout the globe.  However, it is still a client-server solution, and the offending government can easily identify local OSN servers and shut down the links to them.  Instead, we believe a next-generation OSN should be able to function even if the link to the outside world is severed.  Local communication can still be effectual without requiring a connection to the rest of the Internet. Though certainly the social experience is enhanced once the link to the outside world is re-established. Furthermore, this property of resiliency to fragmentation should hold at smaller granularities, so that there should still be islands of connectivity within which users can communicate despite censorship efforts aimed at further partitioning.

There are a variety of OSN services available today and the number of OSN service providers is increasing on daily basis. This abundance of choice (due to features, policies etc.) has caused existing users to be inefficient when it comes to managing their online profiles on different OSNs. In addition, it has caused new users to wander when it comes to selecting an OSN. Using different user interfaces across different OSNs and adapting to confusing changes and features within the same OSN is also inconvenient for many users. Using a single social network service that adapts to the needs of its users replacing all other OSN services is the ultimate goal that would address these shortcomings.

Finally, hosting a huge amount of user information has made OSN service providers an attractive target of security attacks and the fact that a limited set of data centers are hosting all the user information would amplify the damage caused by a successful attack. It is evident that as the demand for OSNs increases, it intensifies the negative effect of these shortcomings  on user experience, and at some point, the need for an alternative approach to todays OSNs is imminent. Hence motivating us to design a new OSN architecture that would address all these shortcomings, while preserving all the benefits of conventional OSNs.

Almost all the obstacles mentioned earlier are somehow caused because of the centralized nature of conventional OSNs. In this technical report, we envision {\it MyZone} a next-generation OSN that moves beyond the centralized architecture of today's conventional OSNs, and instead incorporates distributed P2P networking in a way that is resilient, privacy-preserving, and secure.

The rest of this technical report is organized as follows, next chapter reviews the related works and research on peer to peer social networks, chapter \ref{ch:assumptions} states the assumptions about the system environment where MyZone would be deployed. Chapter \ref{ch:designGoalsAndChallenges} describes challenges that are faced while designing such an OSN service, in addition to the expected functionalities and guarantees that the system should provide.

In chapter \ref{ch:systemDesign}, we describe our two layered design architecture of MyZone. The lower layer provides a secure infrastructure in form of a set of services for the upper layer, which implements the features supported by MyZone. Chapter \ref{ch:securityMeasures} describes the measures embedded into the design in order to address security attacks under different scenarios. Chapter \ref{ch:implementation} describes the packages implemented at the service layer and defines the interfaces for each component of the service layer. Finally, in chapter \ref{ch:conclusion}, we summarize our contributions and identify future research directions.

\section{Related Works}
\label{ch:related}

There have been several attempts at P2P social networks, motivated by the desire to achieve user privacy. In this section we briefly visit some of the existing works while outlining their insufficiencies. Cutillo et al.\cite{safebook} propose a decentralized and privacy-preserving OSN application called Safebook that provides two types of overlays: A set of {\it matryoshkas} which are concentric structures created around each node to provide data storage and communication privacy, and a P2P substrate providing lookup services.  The process of building a matryoshka for a node starts at that node as a {\it core} selecting some {\it trusted} friends as mirrors and creating the first layer. This process will be repeated by each selected friend until a sufficient number of layers is constructed around the core node. Each node in the matryoshka only knows about the identity of its neighbors in the adjacent layers. A message can only be routed to a core node through its matryoshka starting at the outermost layer. Any kind of offline communication can be served by the mirrors. The main purpose of safebook is to prevent identity theft and backtracking of user requests. In the process it also achieves availability by replicating user profiles on their mirrors i.e. the nodes on the innermost layer of user's matryoshka.

There are several problems with this approach. First, a node can only join safebook by invitation since it needs to ask the existing node to create the matryoshka. Another problem is that the reachability of a user on safebook depends on the number of mirrors i.e. first layer nodes, that it can obtain and since this number is fixed obtaining that many nodes is another problem. Furthermore, in case the user is not online, the offline messages like wall posts are stored in encrypted format retrievable only by the user on the mirrors. This means that these updates are not accessible until the user comes back online and publishes those updates. Safebook uses DHTs to find entry points for each node and DHTs are maintained by a third party so the system would be dependent upon another P2P system. Another serious problem with safebook is that it does not address friendship revocation that is an inherent part of any OSN. Finally, using the overlay matryoshkas would result in longer paths on the IP infrastructure yielding in poor performance. 

Buchegger et al.\cite{peerson} proposed another distributed P2P social network application called PeerSoN. PeerSoN has a two-tier architecture: a look up tier that uses DHT and a second tier that consists of peers and contains the user data, such as user profiles. They provide a very high level design of the system and their prototype did not provide encryption at the time. In PeerSoN it is implicitly assumed that the users are connected most of the time although it may be through different devices and from different locations. It is not clear if and how these devices would be synced with each other. Offline messages are stored on the DHT which means that the system can't be scalable since DHT has limited storage. Furthermore, the authors did not address the problem of replication which means that the profile of a user is not accessible if she is offline. Finally, friendship revocation is not considered as a functionality of the system. 

Vis-$\grave{a}$-Vis\cite{vis-a-vis} is an OSN design that uses a Virtual Individual Servers (VIS) as a personal virtual machine running in a paid compute facility. In Vis-$\grave{a}$-Vis, a person stores her data on her own VIS, which arbitrates access to that data by others. VISs self-organize into overlay networks corresponding to social groups. Vis-$\grave{a}$-Vis is designed so that it can interoperate with existing OSNs. Users are responsible to distribute their public keys and the IP address of their VISs. The main problem with this approach is that users have to pay for a VIS and in many cases take charge of maintenance of their own VIS.

Cuckoo\cite{cuckoo} is a microblogging decentralized social network that uses a centralized OSNs only as backup and as a byproduct saving bandwidth costs and reducing downtime while performing equally well. Cuckoo organizes user clients into a structured P2P overlay using pastry. Besides the Pastry routing table, each user also maintains four lists for friends, neighbors, followings and followers. A user maintains connection with $m$ of its online friends and together make up a virtual node $VN$ via request redirection. Each user also maintains $n$ random followers. The idea revolves around the fact that $m$ and $n$ are chosen so that if a user has a small number of friends or followers it would store all of them otherwise it would store a fraction of them. Returning users retrieve updates by using flood-based search for influential users and DHT based search for normal users. Content propagation is done in two ways. A normal users directly pushes messages to his followers while contents generated by influentials are pushed using gossip-based methods between neighbors. Cuckoo does not provide any encryption and can only be applied to microblogging services like Twitter with very limited functionalities. Finally, all the previous works implicitly assume public IP addresses for peers and they don't address the problem of NAT traversal.

A variety of past and ongoing projects on P2P social networking include Diaspora\cite{diaspora}, The Appleseed Project\cite{appleseed}, and Peerbook\cite{peerbook}. One serious shortcoming of these projects is that, they require an actual server that is up an running all the time. Obtaining a publicly accessible server is not free of charge, and it needs regular monitoring and maintenance on behalf of the user, requiring server administration knowledge. In spite of free OSN services, it is very unlikely that these solutions would succeed. 

In addition, there is existing work on secure and anonymous P2P networking, e.g. Darknet\cite{darknet}, GNUnet\cite{gnunet}, I2P\cite{iip}, Bunzilla\cite{bunzilla}, and Freenet\cite{freenet}.  In spite of all these projects, and in spite of all the obstacles facing conventional OSNs, still it is safe to say that P2P social networking is far from being considered as a threat to OSN providers Due to their distributed nature, in large measure because the aforementioned P2P OSNs lack major competing features compared to centralized OSNs.

At the current state, the trend seems to indicate that users would prefer features and convenient user experience at the cost of privacy violations. Therefore, we believe that P2P OSNs would only be able to compete with their centralized counterparts if they can provide the same features and functionalities at the same level of user experience.  MyZone is designed to ultimately achieve the goal of providing competitive features offered by conventional OSNs while supporting key benefits like privacy, security, and resiliency.

\section{Assumptions}
\label{ch:assumptions}

In this chapter we first state the system properties that MyZone requires in order to function. This is followed by the definition of the security model composed of {\bf 1)} a trust model that describes the trusted entities in the system and their trust relations with each other, and {\bf 2)} an adversary model that identifies different adversaries, their resources, and the kind of attacks that they can implement.

\subsection{System Requirements}
\label{sec:systemRequirements}

We divide the system requirements into two categories: participating devices and network infrastructure. We make the following assumption for participating devices:

\begin{enumerate}
\item The devices can be desktops, laptops or smartphones with network connectivity either through ethernet or wireless.
\item Although each device can be behind any kind of firewall or NAT, a reasonably small number of devices should be able to obtain public IP addresses.
\item Out of all those devices with public IP addresses, at least one, should have two network connections, i.e. dual homed, with different public IP addresses. 
\item A very small fraction of those participating devices with public IP addresses e.g. one to ten percent, should be able to host databases that serve a reasonably small number of peers. 
\item Finally, for the purpose of platform independency, we have used Java to implement the services provided to the application. Therefore, in our case, we require each device to be capable of running Java applications. This requirement is not restrictive since Java is installable on all computers and in case of smartphones, Android based smartphones come with this capability.
\end{enumerate}

The network requirements are very simple and include the following:

\begin{enumerate}
\item There exists a functioning IP infrastructure at all times, even though it may be partitioned to some extent either intentionally or due to unpredictable outages. 
\item For enhanced user experience, a minimum bandwidth of 256 Kbps\footnote{Even in many developing countries this bandwidth is widely available to home users.} is recommended but not required. 
\end{enumerate}

\subsection{Security Model}
\label{sec:securityModel}

As mentioned earlier, the security model consists of a trust model, that defines the trusted entities and their relationships with each other, and an adversary model, that identifies different types of adversaries, their capabilities, and the kind of attacks that they can implement. We first start with the description of the trust model.

\subsubsection{Trust Model}
\label{subsubsec:trustmodel}

We define four levels of trust in MyZone:
\begin{itemize}
\item {\it Trust as a certificate authority}: Users put their utmost trust in $CA$ to act as a certificate authority for the social network. The $CA$ issues certificates to all users, and users can verify all the certificates that are issued by the $CA$. 

We assume that the $CA$ is not penetrable by malicious nodes, while it can be target of other attacks. Also, the $CA$ will never act maliciously during its lifetime. These assumptions are essential for the OSN to function, and everything falls apart if one of these assumptions fails.  
\item {\it Trust as a user}: User $A$ can verify the identity of user $B$. This can be done through a certificate issued to $B$ by a certificate authority (CA) that is trusted by $A$.
\item {\it Trust as a friend}: User $A$ trusts user $B$ as her friend and gives $B$ read and write access to its profile. This trust is symmetric but not transitive.
\item {\it Trust as a mirror}: User $A$ trusts user $B$ as her mirror and grants $B$ permission to serve as $A$'s mirror whenever $A$ is unavailable. User $A$ already trusts $B$ as a friend, hence, this is a stronger level of trust.
\item {\it Trust as a replica}: In this scenario, user $A$ is a friend of user $C$ and user $B$ is a mirror for user $C$. User $A$ trusts user $B$ to act as a benevolent mirror on behalf of user $C$ whenever $C$ is unavailable. 

This kind of trust can be referred to as ``{\it trust by proxy}'' and it is in a way, derived based on a transitive relationship on ``{\it trust as a friend}'' and ``{\it trust as a mirror}''. One important property of this type of trust, is one directional read and write access, i.e. only $A$ can read from and write to $C$'s replicated profile on $B$, and $B$ can't modify or read $A$'s profile. 
\end{itemize} 

We assume that two friends will not act as adversaries of each other. This does not apply to friends of friends, since trust is not transitive. We also assume that a benevolent mirror is trusted with the integrity of data (it will not maliciously modify the data) but it may act to gain profile information from its clients, i.e. mirrors can be ``honest but
curious'' users.

\subsubsection{Adversary Model}
\label{subsubsec:adversary}

\noindent{\it Types of adversaries}: There are three types of adversaries
that may be present in the system:
\begin{itemize}
\item Users that are {\it ``honest but curious''} who want to view the user profiles of non-friend users.
\item Malicious entities that are not users of the OSN, but want to attack the system in one of the following ways:
\begin{itemize}
\item Eavesdropping.
\item Spoofing: This includes DNS hijacking. 
\item (Distributed) Denial of Service.
\item IP or URL filtering.
\item Backtracking: an attempt to link some content to a user, e.g. a government agency trying to find the user who is responsible for some particular content.
\end{itemize}
\item Malicious entities that are also users of the OSN. We assume that these entities are not friends of the users they are trying to attack. These entities are perhaps, the most dangerous of all.  They can have resources to implement the following attacks on peer to peer systems, such as our design, that use DHTs:
\begin{itemize}
\item Sybil attacks: an attacker creates a large number of identities and dominates the overlay network, by fooling the protocols, and subverting mechanisms based on redundancy.
\item Eclipse attacks also known as routing table poisoning: an attacker controls a sufficient fraction of the neighbors of correct nodes. Hence some correct nodes can be {\it eclipsed}.
This type of attack applies to network proximity based DHTs such as Pastry
\cite{pastry} and Tapestry\cite{tapestry}.
\item Routing attacks: an attacker may do a combination of the following:
\begin{enumerate} 
\item Refuse to forward a lookup request. 
\item It may forward it to an incorrect, non-existing, or malicious node. 
\item It may pretend to be the node responsible for the key.
\end{enumerate}
\item Storage attacks: a node routes requests correctly, but denies the existence of a valid key or provides invalid data as a response.
\end{itemize}
\end{itemize}

\noindent{\it Motivations of adversaries}: In general, malicious entities have two motivations:
\begin{enumerate}
\item Gather information about users and use it against individuals or the social network as a whole, e.g. implementing a divide and rule policy among users by spreading mistrust.
\item Prevent access to the social network, and hence, create an outage to a subset of the system.
\end{enumerate}

\noindent{\it Adversary resources}: We make the following assumptions in
terms of the resources available to an adversary:
\begin{itemize}
\item An adversary has access to {\it ``portions''} of the IP infrastructure and can filter the URL or IP address of individual devices on the network. 
\item The number of IP addresses and URLs that an adversary can filter is limited to only a small fraction of the number of overall devices that participate in the social network. 
\item An adversary has the ability to execute a successful DDOS attack on a limited set of devices at any point in time.
\item An adversary does not have sufficient computational power to crack strong symmetric or asymmetric cryptographies.
\item Finally, an adversary has enough resources to create a limited number of malicious entities in different roles as components of the OSN. 
\end{itemize}

\section{Design Goals and Challenges}
\label{ch:designGoalsAndChallenges}

In this chapter we describe the main goals that are motivating the design of MyZone, and describe the challenges facing it.
 
\subsection{Design Goals}
\label{sec:goals}

In chapter \ref{ch:intro} we described the following shortcomings for the current OSNs:

\begin{enumerate}
\item User privacy violations on behalf of the OSN service providers.
\item Government imposed censorship aimed at repressing popular uprisings by means of preventing the OSN services being used as tools for facilitating these movements.  
\item User confusion and frustration resulting from the abundance of different OSNs, and the frequently changing policies and features that are imposed by the OSN service providers.
\item The undeniable threat of hijacking user information from centralized data centers by targeted attacks.  
\end{enumerate} 

The eminent need for a next generation of OSNs that addresses these shortcomings has motivated MyZone towards the following goals:

\begin{itemize}

\item It should preserve user privacy by ensuring that users own their own data, and no central authority or third party can access user information without explicit permission from the user.  

\item User privacy preservation should not come at the price of losing functionalities of conventional OSNs. Therefore, it should support all primary functionalities of a conventional OSN like Facebook. These include the ability to establish and revoke friendships, sharing contents and comments with friends at different levels, and participating in discussions.

\item It should be highly resilient in presence of malicious attacks as well as infrastructure failures caused by unpredictable causes. The OSN continues to function in isolated islands (which will be referred to as {\it ``local deployment''}) while facing network partitioning, and possesses self healing properties, by automatically reconciling the states after the partition has been restored.

\item It should tolerate a high rate of churn due to mobile users, and seek to provide
near 24/7 availability of user profiles (best effort).

\item Due to the growing trend of switching to smartphones and tablet computers, the underlying computing platform consists of traditional desktops and laptops with high bandwidth network connections, as well as portable devices with restricted resources. To deal with the latter, the system should be designed to be minimally demanding of the limited resources of portable devices, and should incorporate power/resource management techniques.

\end{itemize}

\subsection{Challenges}

The shortcomings of conventional OSNs, demand a distributed architecture for the next generation of OSNs. Based on the distributed architecture, the assumptions mentioned earlier, and the design goals, we identify six key challenges in the design, implementation, and deployment of
MyZone: availability, resiliency, routing, connectivity, security, and traffic optimization and power management.

\subsubsection{Availability Challenges}
\label{subsec:availability}

An important property of OSNs is that a user profile is always available irrespective of whether that user is currently online or not. This is made possible by storing user profiles in a central server that is available all the time. Providing this property in a distributed OSN, where user profiles are not replicated at any central server is further complicated by high churn rate. Overcoming this challenge is perhaps the key to the success of a distributed OSN. 

In the absence of a central server, profile replication must be considered as an intrinsic part of the system design. The key question is where should a user profile be replicated, i.e. how to select {\bf mirrors} for replicating a user profile. There are two design choices with this regard:
mirroring a user profile on a set of devices belonging to unfamiliar users; or replicating the user profile on a set of devices belonging to users that are trusted by the owner of the profile.

Employing DHT services would use the first approach to replicate user profiles. Although DHTs have proven to be a success, dealing with availability issues, they are mostly used to replicate contents that are not going to be modified. Even in some cases where the replicated content is modified, the modifying entities are not restricted to a specific selection of users. One possibility of implementing this restriction is by sharing a common key between all friends of a user and replicate the encrypted profiles, using this common key. 

There are three problems with this approach. First, even assuming that the shared keys are not going to be obtained by non-friend users, revoking a friendship translates into revoking the shared key, issuing a new key, sending it to all friends, and finally re-encrypting the entire profile using the new key. This process is very inefficient considering the fact that revoking friendships can occur frequently. 

Second, a simple modification of the user profile by a friend e.g. a one word comment, may translate into sending the entire encrypted profile to the friend and re-encrypting it by the friend after making the modification, and sending it back to the mirror. This process is also inefficient and would result in unnecessary use of bandwidth, traffic, and power resources. 

Even if the updates are encrypted in a way that prevents this to happen e.g. sending encrypted updates to the mirror and appending them to the end of the encrypted profile, it is still very hard to send a particular part of the profile from the mirror to the friend since the entire profile is encrypted and the mirror does not know which chunks of the encrypted profile should be sent.

Finally and probably most importantly it would be impossible to implement application level permissions e.g. who can see what part of profile or modify it, since the profile is encrypted using a {\it shared} key. Furthermore, the mirror will not be able to impose those permissions since it does not know the key. Of course sharing the key with an unknown mirror would compromise the user privacy and profile integrity. Thus, because of the special features and functionalities of OSNs, DHTs can not and should not be used to host OSN services.

The only feasible solution would be the second approach and we believe that candidate designs must investigate the second approach, where users find other trustworthy users amongst existing friends to replicate their profiles. Finding friends who are willing to be mirrors and convincing them  to serve as mirror either through social ties, or proposing incentives, would be an interesting problem on its own.

In addition to the issue of selecting mirrors, there are several other aspects of replication management that must be addressed given the dynamic nature of the system due to churn. These  are: 
\begin{enumerate}
\item When and how profile updates are propagated to other mirrors.
\item How to maintain replica consistency among all mirrors.
\item What type of replica consistency is appropriate for this application.
\item Whether a user profile is completely or partially replicated at different mirrors.
\item Whether the mirrors themselves should be prioritized as primary, secondary,  and so on. 
\end{enumerate}

\subsubsection{Resiliency Challenges}
The proposed system is intended to be deployed in an infrastructure that is vulnerable due to: {\bf 1)} mobile users with widely varying Internet connections,  and {\bf 2)} malicious entities that may try to partition the social network or bring down the system. Hence the system needs to be resilient in such environments and must be self-healing.

\subsubsection{Routing Challenges}

A key requirement of the OSN is that, users must be able to locate their (mobile) friends and establish connections when they join the system. This is further complicated by the high rate of churn.

One approach is that all users register their IP addresses on a rendezvous server when they join the system . A user then, would contact this server
with appropriate credentials to determine the current IP address of her friend. This solution suffers from the major drawback of the server being a single point of failure.  In addition, that single rendezvous server needs to be trusted since it would have a global view of the social graph and it can derive social relationships, which is not desirable. 

A second approach is to use a distributed hash table (DHT) scheme to hash an individual's username to a particular rendezvous server for lookup, but then we have to implement an efficient mechanism that would prevent the rendezvous server in charge of a user $i$ from disclosing user $i$'s IP address to users that are not friends with $i$. This is to prevent targeted attacks on $i$. Our design would focus on the second approach.

\subsubsection{Connectivity Challenges}

The user devices participating in the OSN are usually (around 90\%) behind firewalls or NATs. NATs are divided into four types: 
\begin{itemize}
\item Full Cone NAT: Once an internal address (iAddr:iPort) is mapped to an external address (eAddr:ePort), any packets from iAddr:iPort will be sent through eAddr:ePort and any external host can send packets to iAddr:iPort by sending packets to eAddr:ePort.
\item Address Restricted Cone NAT: Once an internal address (iAddr:iPort) is mapped to an external address (eAddr:ePort), any packets from iAddr:iPort will be sent through eAddr:ePort and an external host (hAddr:any) can send packets to iAddr:iPort by sending packets to eAddr:ePort only if iAddr:iPort has previously sent a packet to hAddr:any. "Any" means the port number doesn't matter.
\item Port Restricted Cone NAT: Once an internal address (iAddr:iPort) is mapped to an external address (eAddr:ePort), any packets from iAddr:iPort will be sent through eAddr:ePort and an external host (hAddr:hPort) can send packets to iAddr:iPort by sending packets to eAddr:ePort only if iAddr:iPort has previously sent a packet to hAddr:hPort.
\item Symmetric Cone NAT: The NAT mapping refers specifically to the connection between the local host address and port number and the destination address and port number and a binding of the local address and port to a public side address and port. Any attempts to change any one of these fields requires a different NAT binding. 

This is the most restrictive form of NAT behavior under UDP, and it has been observed that this form of NAT behavior is becoming quite rare, because it prevents the operation of all forms of applications that undertake referral and handover.
\end{itemize} 

Configuring firewalls and NATs to forward requests to a particular internal host is not common knowledge for many users, and may not be possible in some cases. An alternative way to avoid reconfiguring NAT is called NAT traversal. 

A common method of NAT traversal is for an internal host to send a UDP packet to an outside host first, and depending on the type of NAT, the internal host may be able to receive connections from all, or a subset of external hosts afterwards. This method is commonly known as UDP hole punching. Except for the full cone NAT, the use of UDP hole punching limits the extent of external hosts that can connect to the internal host. In fact UDP hole punching is impossible on symmetric cone NATs. 

Although full cone NATs are commonly used by home users, a good portion of users are behind other kinds of NAT. Connecting peers that are behind NATs is another challenge that must be addressed, otherwise the usage of OSN would be limited to very few users that have access to public IP addresses.

\subsubsection{Security Challenges}

Our design needs to provide the following security guarantees:

\begin{itemize}
\item {\it Confidentiality}: Prevent the disclosure of profile information to any user that is not a friend.
\item {\it Integrity}: A user profile can only be viewed and modified by her friends.
\item {\it Availability}: The OSN is resilient against DDoS attacks and URL/IP filtering, and user profiles are {\it ``almost always''} available.
\item {\it Authenticity}: Every modification of a user profile is authenticated.
\item {\it Consistency}: A weakly consistent view of the profile is always available.
\end{itemize}

The consistency and availability guarantees are bound to be best effort as the benefits gained through absolute availability and total consistency are not worth the added complexity in the design of the OSN.  This tradeoff is described in the CAP Theorem\cite{brewer2000towards}, which stipulates that there is a tradeoff in distributed systems between consistency, availability, and partitioning (in our case due to attacks and/or mobile churn).

\subsubsection{Traffic Optimization and Power Management Challenges}
\label{subsec:traffic_power}

As mentioned in chapter \ref{ch:assumptions}, users may use a wide range of devices to participate on the OSN, including smartphones and other portable devices, laptops and desktop computers. In addition each user device may also act as a mirror for another user. A key challenge is to provide availability while dealing with resource-constrained devices.  Portable devices are primarily constrained by energy and bandwidth limitations, though memory and CPU limitations also play a factor.  A challenge is to develop a replication strategy that takes into account these limitations while providing availability and resiliency. One strategy is to leverage more resource-rich devices, i.e. online desktops, to support the bulk of mirroring duties, and to use mobile mirroring only as a strategy of last resort.

Heterogeneity of device resources and their limitations call for optimized resource usage of the system, especially with regards to traffic and power management schemes. Such a system should try to save as much energy and bandwidth as possible (memory and CPU are not as critical as power and bandwidth in portable devices). For example, a pushing scheme would prove less efficient in terms of power and traffic optimization as opposed to a pulling scheme when it comes to reflecting user profile updates to friends.

\section{System Design Architecture}
\label{ch:systemDesign}

This chapter gives a detailed description of our design. Our design is based on a two layered architecture. The lower layer referred to as the {\it service layer} provides essential services to the components of the system and facilitates the registration of peers, finding peers, establishing connections between peers and much more. 

These services are designed to provide a reliable, resilient and secure infrastructure addressing the requirements of peer to peer applications, from NAT traversal, to reliable UDP connections, to secure socket services, and more. Hence, the service layer can be used by any peer to peer application that requires these features and is not specific to just OSNs. 

The upper layer is referred to as the {\it application layer}. This layer provides the specific features and functionalities of the OSN. In addition, it is responsible for implementing higher level security policies such as read and write permissions for particular users or groups of users. Profile replication is also done at this layer. Next we describe the two layered architecture starting from the lower layer.

\subsection{Service Layer}
\label{sec:serviceLayer}

The service layer provides an infrastructure that is resilient against partitioning, reliable even on top of UDP, and secure against malicious attacks. It is also very dynamic and can be used by peers behind all kinds of firewalls and NATs. In this section we describe the service layer by describing its components and their interactions with each other. 

We divide our proposed design of the service layer into two parts based on the nature of their deployment. A local service layer that is used for {\it local deployment}, and a global service layer that is used for {\it global deployment}. The purpose of local deployment is to establish a private OSN intended to be used in hostile environments. 

Perhaps a very useful application of a private OSN is in democratic movements opposed by the government entities, where the OSN services can be used to organize protests, spread news, share ideas, and much more, when the access to the outside world is blocked or limited. 

In fact the local deployment model can be used in form of a package that we call {\it Democracy IN A Box (Dinab)} for these specific purposes. To assure maximum security, our entire design builds upon {\it ``the need to know basis''} philosophy which mandates that any information would only be available to those entities that {\it must} know about it. 

The local deployment is intended to be used by a small number of users, compared to the global deployment where the number of users is unbounded. The use of OSN by the general public is considered as global deployment. In our design the global deployment model is built on top of the local deployment model. Therefore, we describe the local deployment model first, and then we explain how the global deployment model builds on top of it.

Let's examine the steps that a peer should take, in order to establish a secure connection with another peer. A peer can directly connect to another peer if, 1) the IP address of the other peer is known, and 2) the peer either has a public IP address, or is behind a NAT that can be traversed using UDP hole punching. If the NAT traversal is not possible then the peer can only receive connections through a relay. 

In order to be able to establish a secure connection between peers, they should be able to verify the identities of each other. This can be done by using certificates, which means, a certificate authority that is trusted by all peers needs to exist. Based on the steps mentioned earlier, the local deployment model consists of five components:
\begin{itemize}
\item STUN server: STUN ({\it Simple Traversal of User datagram protocol (UDP) through Network address translators}) features an algorithm to allow endpoints to determine NAT behaviors. The STUN server is needed so that peers can determine the type of NAT they are behind. The protocol is lightweight and is documented in RFC 3489. The STUN server is the only entity that needs to run on a dual homed machine with two public IP addresses. 

\item Certificate Authority: The CA server would issue certificates to each new user and verifies that the usernames are unique. We assume that peers have already obtained the public key of the CA which is needed to verify the certificates. Note that several CAs can be available in the system and a user can have several certificates issued by different CAs. 
\item Rendezvous Server: Peers need to register with rendezvous servers in order to be found by other peers. The server should be able to host a database. This database stores the information for peers as well as relays. In case, a peer needs a relay to accept connections, it will query the rendezvous server for a relay server. 
\item Relay Server: The relay server would relay the connection between two peers in scenarios, where the peer acting as server is behind an un-traversable NAT. The relay server can't decrypt the relayed connection and therefore, the relayed connection is viewed as an end to end secure connection between the two peers.
\item Peers: The devices that host the user profiles and represent users. 
\end{itemize}

\begin{figure*} \begin{center} \scalebox{0.33}{\includegraphics{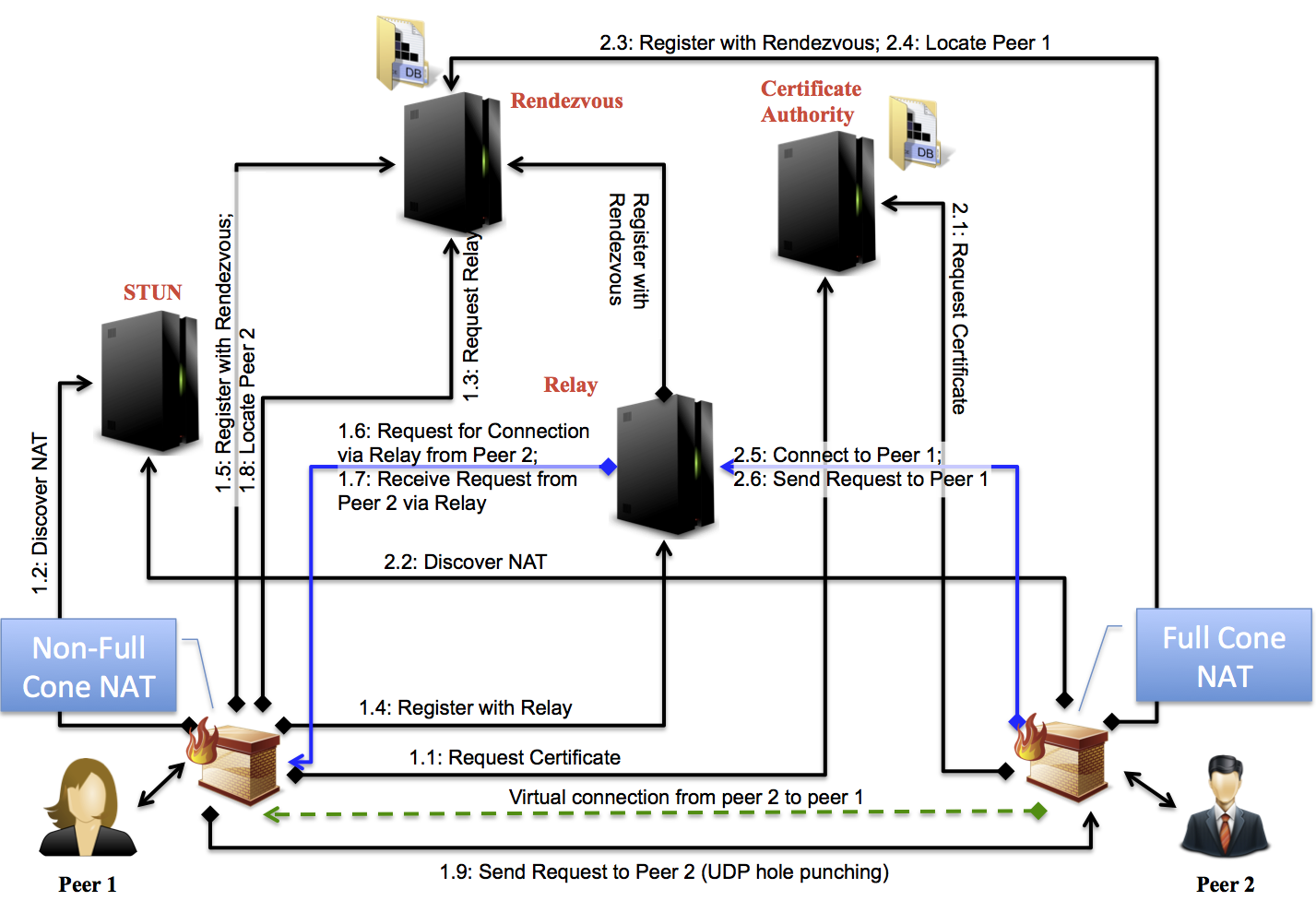}} \end{center}
\caption{Deployment diagram for local deployment  . \label{local_deployment} }
\end{figure*}

Figure \ref{local_deployment} shows the components of the service layer for local deployment and the sequence of steps to establish a connection between two peers. The sequence of steps for each peer is represented as $i.x$ where $i$ is the peer and $x$ is the step.

\begin{figure*} \begin{center} \scalebox{0.53}{\includegraphics{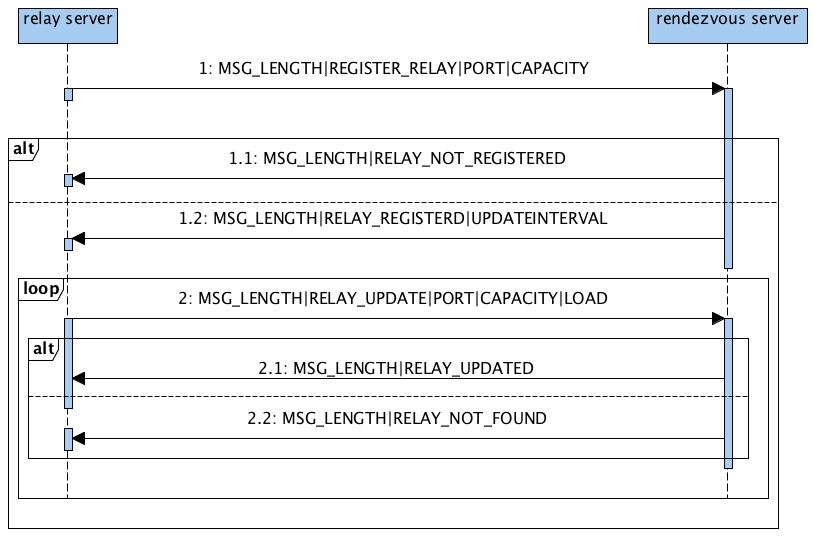}} \end{center}
\caption{Sequence digram of the scenario where a relay server registers with a rendezvous server. \label{relay_registration_sd} }
\end{figure*}

Now we describe the interactions between the components, based on the sequence of steps in Figure \ref{local_deployment}. Prior to the steps that a peer should take, a relay server must register with a rendezvous server so that it can be reached by peers that are behind un-traversable NATs. Figure \ref{relay_registration_sd} describes the sequence of messages exchanged between the relay server and the rendezvous server to achieve this. 

The registration starts by the relay server sending its serving port, and the total number of peers acting as servers that can be relayed, as its capacity. The rendezvous server notifies the relay server if the registration was unsuccessful, or sends back an acknowledgment with a number, indicating the interval in milliseconds, between update packets needed to be sent from the relay server to the rendezvous server. The rendezvous server treats the received update packets, as indicators that the relay server is alive. If the rendezvous server does not receive these update packets for a defined period of time, it will remove the corresponding relay server from its relay server table.

Upon successful registration, the relay server would periodically send out update packets with its current load, capacity, and port number, to indicate that it is alive. 

\begin{figure*} \begin{center} \scalebox{0.4}{\includegraphics{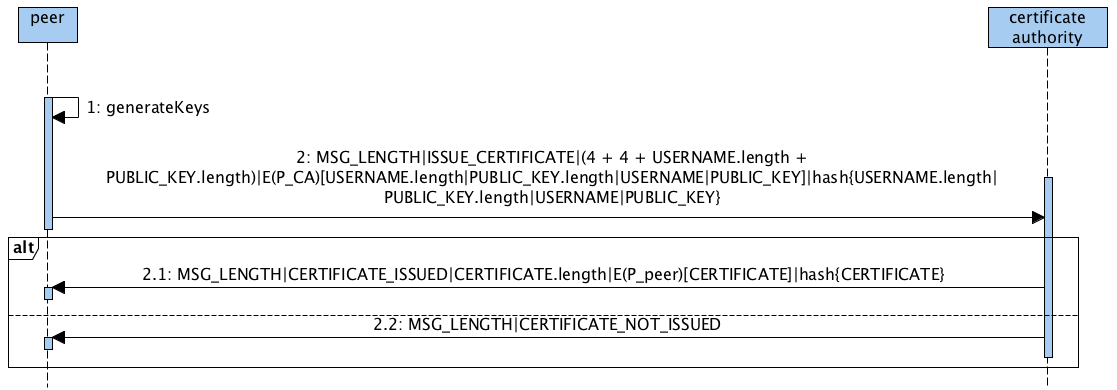}} \end{center}
\caption{Sequence diagram of the scenario where a peer requests a certificate from a certificate authority (CA). \label{request_certificate_sd} }
\end{figure*}

The first step that a peer should take is to obtain a certificate from a trusted certificate authority (CA). This is essential, since a peer needs to be able to prove its identity to other components when required. Figure \ref{request_certificate_sd} shows the sequence of messages exchanged between a peer and the certificate authority (CA) during this process. We remind the reader, that the public key of the CA is already stored on the peers' devices. The peer starts by generating its own pair of keys. It would then send its username and public key, encrypted by an asymmetric encryption algorithm e.g. RSA, using the public key of the CA. 

This message also includes the length of the unencrypted username and public key. This is because, asymmetric encryption uses block cipher, which means that the plain message has to be padded to fill appropriate number of complete blocks before encryption. At the time of decryption the length of the plain message needs to be known, so that the padded part can be trimmed from the decrypted message.

In addition, the message digest will be computed over all plain data that are encrypted, and will be appended to the message. If everything checks out and the username does not already exist on the CA's database, the CA would reply with the issued certificate for the user, encrypted using the user's public key. 

Although encrypting the certificate itself does seem unnecessary at first, it is essential to keep the identity of the requester, hidden from any malicious entity that monitors the requests sent to the CA. This is especially important in hostile environments where a government entity is trying to find the identity of the users that are participating on the OSN by monitoring all the connections going to the CA, as the IP address of the CA is publicly available. 

Note that this step is only done once, and after the initial sign up process, the peer would not need to communicate with the CA anymore. Finally, we emphasize that the CA is the {\it ``only''} entity in the entire design that is assumed to be trusted by all other components and in fact, all other components are assumed to be untrustworthy. 

After obtaining the certificate the next step is for the peer to discover whether it is behind a NAT or firewall and if it is, what type. This is done using the standard STUN protocol mentioned earlier. The reader can refer to RFC 3489 for in depth description of the protocol. If the peer is behind a non-full cone NAT, it would need to obtain a relay server address, and register with it, in order to accept connections from external hosts.

\begin{figure*} \begin{center} \scalebox{0.45}{\includegraphics{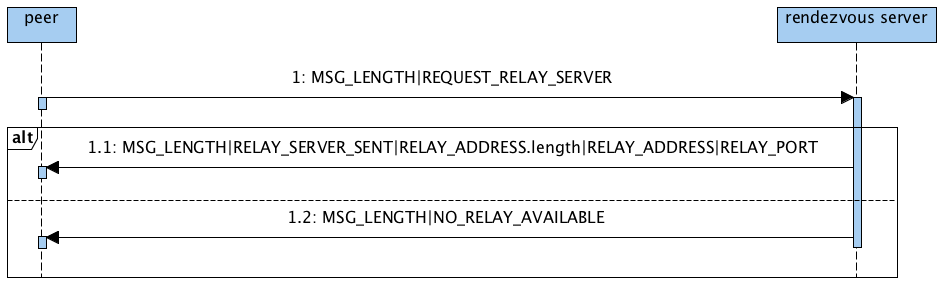}} \end{center}
\caption{Sequence diagram of the scenario where a peer requests for a relay server from a rendezvous server.  \label{request_relay_sd} }
\end{figure*}

Figure \ref{request_relay_sd} shows the process of requesting a relay server address from the rendezvous server. This process is very simple and starts by a peer sending a {\tt request\_relay\_server} message to the rendezvous server and ends by the rendezvous server sending back either the IP address and port number of a relay server, or a {\tt no\_relay\_available} message.

In case there are more than one relay servers registered with the rendezvous server, the rendezvous server selects the relay in a way that ensures balanced loads across all relays. Requesting a relay server does not contain any critical information and can be initiated by anyone even non-participating devices. Hence, the messages exchanged during this process don't need to be encrypted.  
 
\begin{figure*} \begin{center} \scalebox{0.57}{\includegraphics{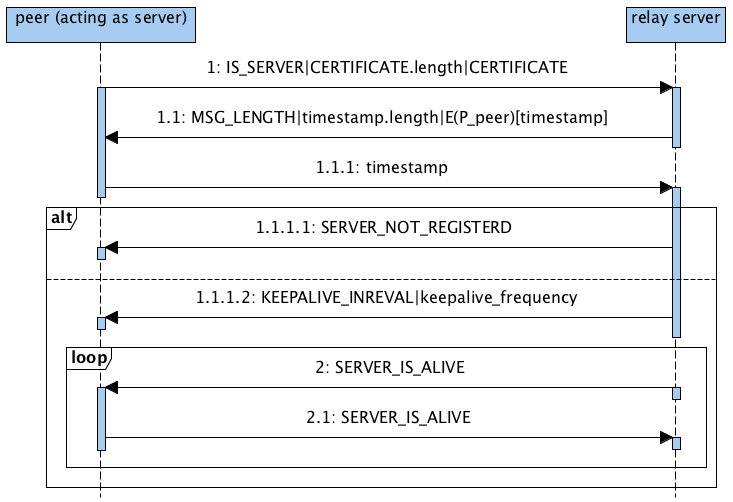}} \end{center}
\caption{Sequence diagram of the scenario where a peer registers with a relay server. \label{peer_registration_with_relay_sd} }
\end{figure*}

After requesting for a relay server address, the peer proceeds by registering with that relay server. Each relay server has to verify the identity of peers that are registering with it. As indicated in Figure \ref{peer_registration_with_relay_sd}, this starts with the peer that is going to receive connections via the relay server, sending its certificate appended to a {\tt is\_server} message, indicating that it is acting as a server. 

The relay would verify the identity of the peer by sending back a timestamp encrypted using the public key extracted from the user's certificate. This timestamp is then decrypted by the user and sent back to the relay server. This method ensures the authenticity of the user and is immune to replay attacks but not man in the middle attacks. We will address man in the middle and other type of attacks in chapter \ref{ch:securityMeasures}. Upon successful registration, a keep alive interval is sent back to the peer. This requires the peer acting as server, to send a {\tt server\_is\_alive} packet to the relay server indicating that it is still alive. This mechanism removes the dead connections from falsely filling the capacity.

\begin{sidewaysfigure*}
\begin{center} \scalebox{0.4}{\includegraphics{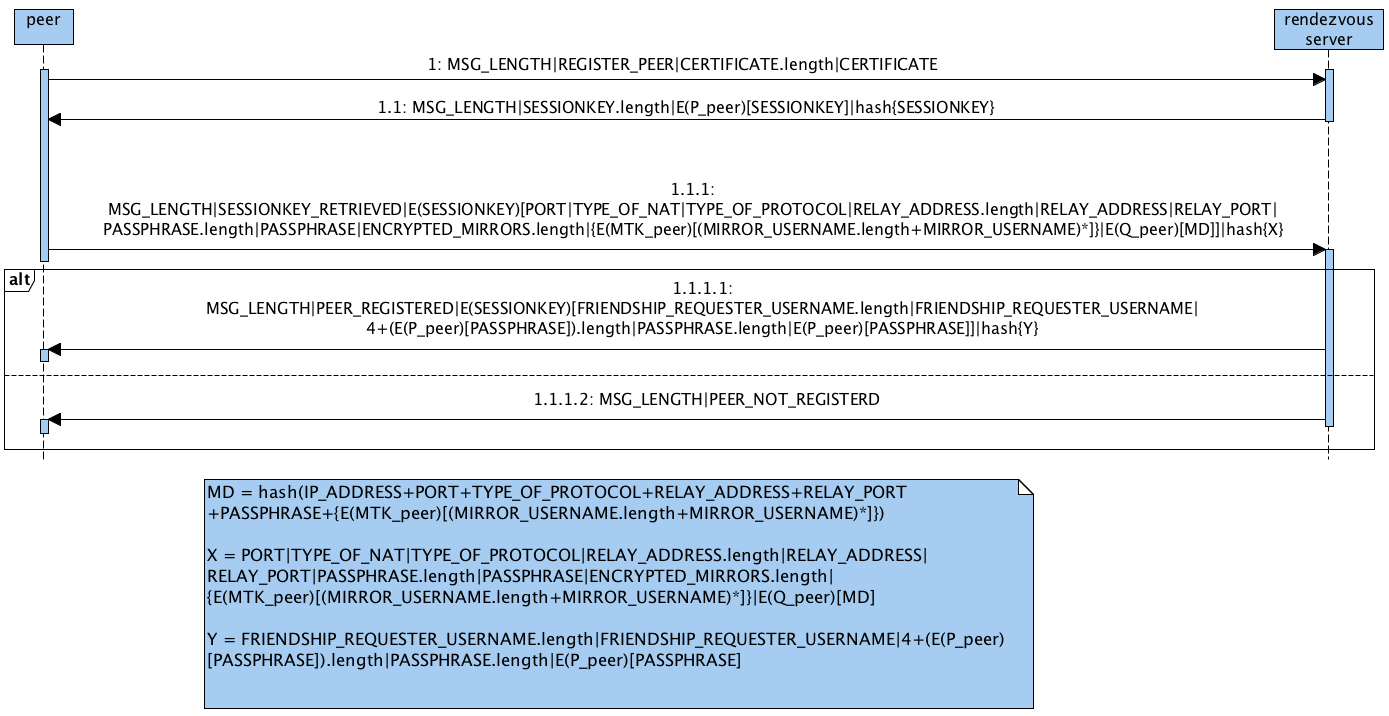}} \end{center}
\caption{Sequence diagram of the scenario where a peer registers with a rendezvous server. \label{peer_registration_sd} }
\end{sidewaysfigure*}

Then comes the important step of a peer, registering with the rendezvous server. Figure \ref{peer_registration_sd} represents the sequence of messages exchanged in this process. The first message is sent by the peer and includes the peer's certificate, while the reply carries the session key generated by the rendezvous server, and encrypted by the public key of the peer. If the session key is correctly retrieved by the peer, it sends the following encrypted information to the rendezvous server:
\begin{enumerate}
\item port number.
\item type of NAT: full cone, non-full cone, or public IP address.
\item type of protocol: TCP or UDP.
\item relay server address.
\item relay server port.
\item passphrase: this is a unique string of characters generated by the peer and is shared with all friends. The passphrase is generated in such a way that it does not hold any relationship with the actual username. 

The look up queries for each peer include the passphrase instead of the actual username, as will be described later. This is to ensure that the access to a user registration information is limited to those who know its passphrase, and not to the general public.
\item encrypted list of mirrors: the mirror list of a peer includes the usernames and the corresponding passphrases of its mirrors. This list is encrypted using a special key generated by the peer, that is shared with all its friends. The encryption would prevent unauthorized entities discovering the identities of mirrors of a particular user.  This complies with our need to know basis design philosophy. 
 \item signed message digest over the following information: IP address, port number, type of protocol, relay address, relay port, passphrase, and the encrypted mirror list. This signed message digest ensures the authenticity and the integrity of the data sent back by the rendezvous server. 
\end{enumerate}  

The rendezvous server will store all this information in its {\it peers} table. Upon successful registration with the rendezvous server, the server returns a {\tt peer\_registered} reply along with optional list of pending friendship requests. The friendship requests for peer $i$ are stored as friendship requester's username, and its passphrase, encrypted using the public key of peer $i$. Encrypting the friendship requester's passphrase is mandated based on the need to know basis philosophy, and prevents malicious entities including rendezvous servers, to look up the IP address of the friendship requester by providing the passphrase. 

\begin{sidewaysfigure*}
\begin{center} \scalebox{0.4}{\includegraphics{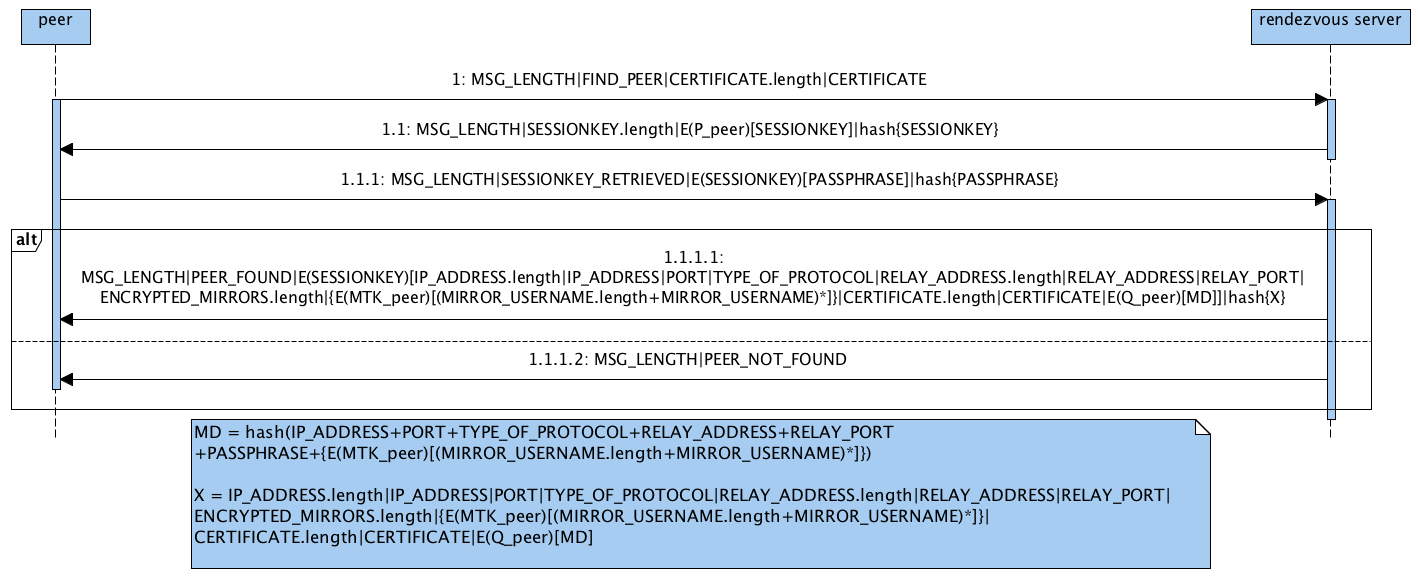}} \end{center}
\caption{Sequence diagram of the scenario where a peer locates another peer. \label{locate_peer_sd} }
\end{sidewaysfigure*}

Peer $i$ can look up the connection information for peer $j$, if 1) peer $j$ has already registered with a rendezvous server and, 2) peer $i$ knows peer $j$'s passphrase. Figure \ref{locate_peer_sd} demonstrates how a look up query is performed. The first two messages establish a secure connection. After the session key is retrieved, the peer sends the passphrase of the target peer to the rendezvous server. Upon finding a username corresponding to that passphrase in its {\it passphrase} table, the rendezvous server replies with the connection information of the target peer. The information includes the signed message digest described earlier, which will be used by the peer to verify the integrity and authenticity of the reply.

\begin{figure*} \begin{center} \scalebox{0.43}{\includegraphics{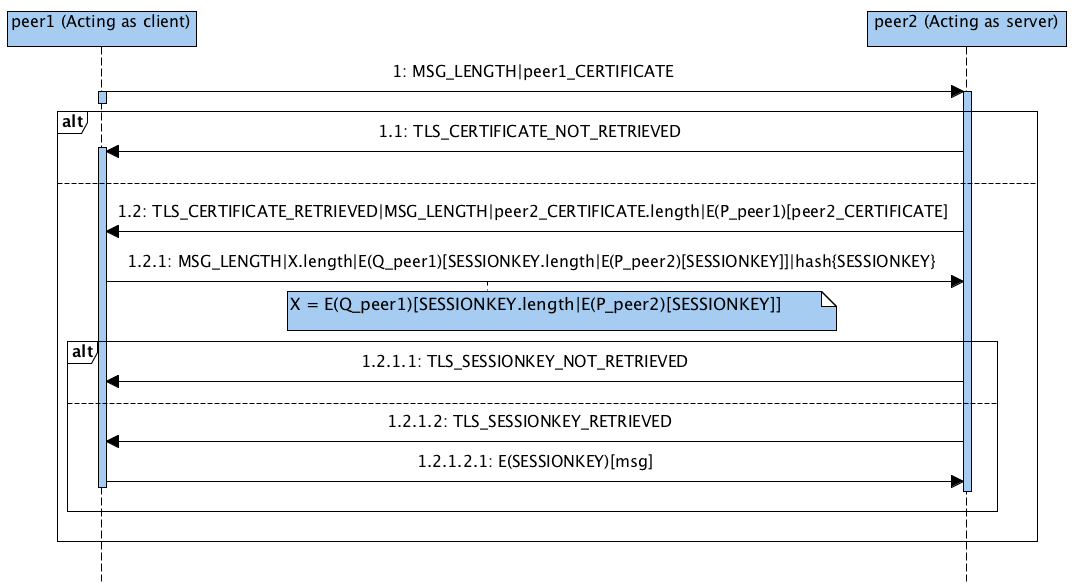}} \end{center}
\caption{Sequence diagram of a scenario where a peer connects to another peer. \label{connecting_peers_sd} }
\end{figure*}

Finally , peer $1$ can connect to peer $2$ after obtaining its connection information using the sequence of messages described in Figure \ref{connecting_peers_sd}. This process starts by peer $1$ sending its certificate to peer $2$. Peer $2$ would then send back its certificate encrypted using peer $1$'s public key. This encryption prevents an entity monitoring peer $1$'s traffic from figuring out the identity of peer $2$.

After the successful exchange of certificates, peer $1$ generates a session key and encrypts it first using peer $2$'s public key and then, its own private key. This double encryption ensures that the session key can only be retrieved by peer $2$ preventing any kind of attack compromising the session key. This single message has a crucial role in the security of the entire service layer. Peer $1$ can securely communicate with peer $2$ after the session key has been successfully retrieved by peer $2$.

\begin{figure*} \begin{center} \scalebox{0.48}{\includegraphics{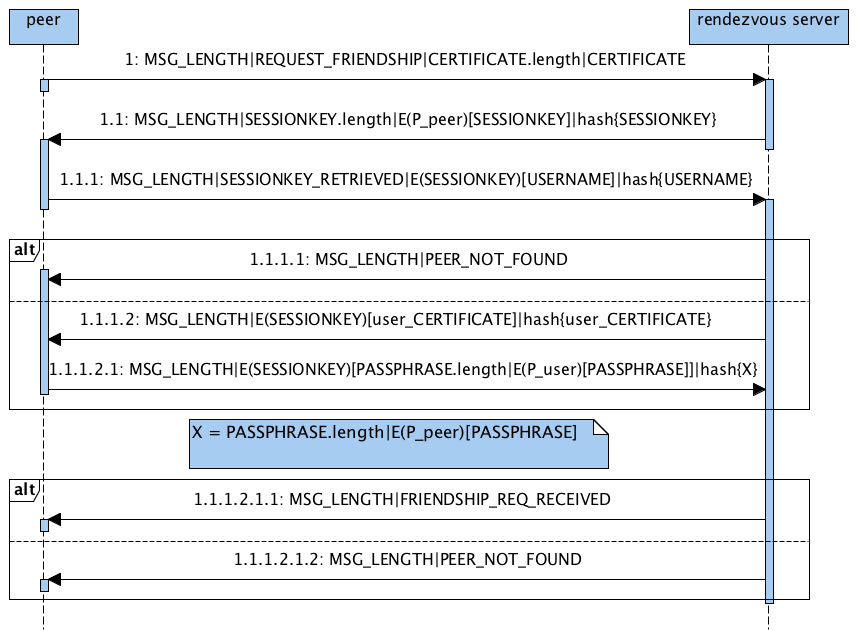}} \end{center}
\caption{Sequence diagram of the scenario where a peer sends friendship request to a rendezvous server. \label{send_friendship_request_sd} }
\end{figure*}

In addition to all previous steps, the service layer also provides a method for a peer to send friendship requests for another peer. All friendship requests for a particular peer $x$ are sent to the rendezvous server responsible for $x$, instead of sending them directly to $x$. There are two reasons for this. First, peer $x$ might not be online at the time of sending the friendship request and therefore may not receive it. Second, all friendship requests not necessarily end in friendships and the connection information of peer $x$ should not be reveled to other peers unless $x$ has accepted their request and has shared its passphrase with them.

Figure \ref{send_friendship_request_sd} shows the sequence of messages exchanged between peer $y$ sending a friendship request for $x$, and the rendezvous server in charge of $x$. This process begins by sending a session key generated by the rendezvous server in order to establish a secure connection. Next, peer $y$ sends the username of $x$ to the rendezvous server. The rendezvous server then, sends back the certificate of $x$, if an entry for $x$ can be found in the peers table of the rendezvous server.

Finally, $y$ sends its passphrase encrypted by the public key of $x$ back to the rendezvous server. The passphrase can only be decrypted by $x$, and therefore, $x$ would be able to look up $y$ if it chooses to accept the friendship request. Friendship revocation at the service layer is done by changing the passphrase and sending the new passphrase to all existing friends so that the deleted user can not access the registration information stored on the rendezvous server. In addition, the secure socket is implemented so that it would only allow access to the application layer if the connected user is a friend, a friend requester, or a requested friend. When a friend is deleted from the friend list the service layer denies all access from that user.

\begin{figure} \begin{center} \scalebox{0.7}{\includegraphics{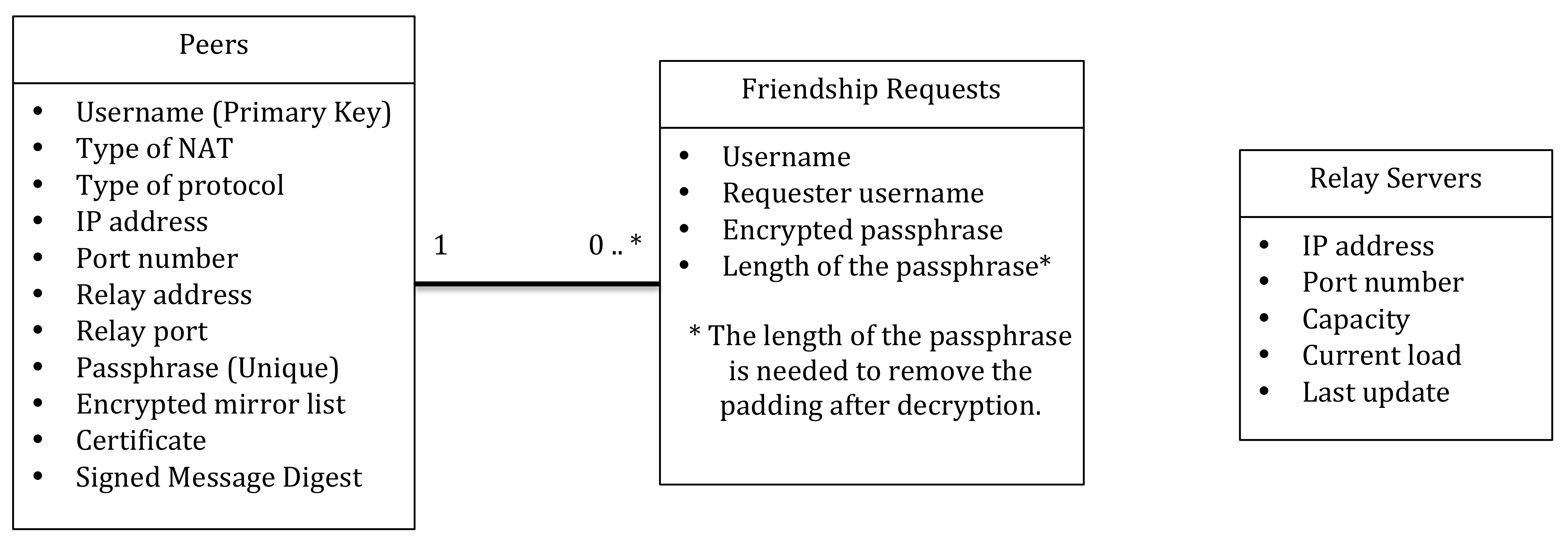}} \end{center}
\caption{Database schema for the rendezvous server database. \label{rendezvous_db} }
\end{figure}

Based on queries that are sent to the rendezvous server during different steps, the rendezvous server needs to host a database. Figure \ref{rendezvous_db} represents the schema of this database. The database includes three different tables: peers, friendship requests, and relay servers. 

The peers table stores all the connection information of registered peers. The friendship requests table stores all the friendship requests for registered peers, and finally, the relay servers table stores all the information of registered relays.

Now that we have described the local deployment service layer, we can introduce the global deployment as an extension of the local deployment model. As mentioned previously in section \ref{subsec:availability} due to the inherent nature of OSNs, DHTs can not be used as stand alone solutions for a large scale deployment. Furthermore, there are several issues that prevent the local deployment model to be used at large scales.

One of these issues is that, although the single rendezvous server can not have access to any of the user profiles, it would be able to compute the social graph and derive all the relationships between users. This violates our need to know basis approach. More importantly, the single rendezvous server would be a huge vulnerability since the entire OSN will be dysfunctional if it goes down. Finally, unauthorized access to the rendezvous server's database can potentially compromise user securities as the attacker would have complete access to the user connection information. 

These issues motivate our design to employ a structured peer to peer system comprising of only the rendezvous servers using chord DHT\cite{chord}. The main reason that we are using chord is because of its simple implementation, its reasonably good performance compared to the other solutions\cite{dht_performance} and some of its inherent features that will be used to prevent different attacks as will be described in chapter \ref{ch:securityMeasures}.

Before we describe our extended design we give a brief description of chord DHT and its properties. Chord uses an overlay network that is shaped as a ring. Chord is designed {\it only} to lookup a {\it key} and return the {\it value} that is responsible for the key. The key can be a filename that is stored on a machine and the value would be the IP address of the machine that hosts the filename. In our case, the key is the username of the peer and the value is the IP address of the rendezvous server that the user should register with.

The key and the value are both mapped into the same domain by use of {\it consistent hashing} ($CH$) and the hashed values are called {\it identifiers}. Consistent hashing\cite{consistent_hashing} has several good properties: {\bf 1)} With high probability the hash function balances load (all nodes receive roughly the same number of keys) also, {\bf 2)} with high probability,
when an {\it N'th} node joins (or leaves) the network, only an $O(1/N)$ fraction of the keys are moved to a different location. This is clearly the minimum necessary to maintain a balanced load.

The chord ring is structured so that a username {\it X} is registered on the rendezvous server {\it Y} if and only if:
$CH(X) \leq CH(Y)$ and there are no rendezvous servers $Z$ where $CH(X) \leq CH(Z) \leq CH(Y)$. Node $Y$ is referred to as {\it Successor(X)}.

Chord has the following properties: 
\begin{itemize}
\item The number of messages that need to be sent to maintain the correct routing table upon leaving or joining the ring is $O(log^2N)$.
\item Routing table at each node only needs to obtain information about $O(logN)$ other nodes on the DHT ring.
\item A successful look up query can be done in $O(logN)$ number of messages. 
\end{itemize}

We are using two types of consistent hashing namely, {\it SHA-1}\cite{sha1} and {\it MD5}\cite{md5}, that would map to the same space. This approach facilitates rendezvous server replication and detection of malicious rendezvous servers as will be explained in chapter \ref{ch:securityMeasures}. We refer the reader to \cite{chord} for complete description of how nodes join a chord ring, replicate data and maintain the correct routing tables. We adapt the same procedures to construct and maintain a chord ring composed of rendezvous servers as participating nodes.

\begin{figure*} \begin{center} \scalebox{0.37}{\includegraphics{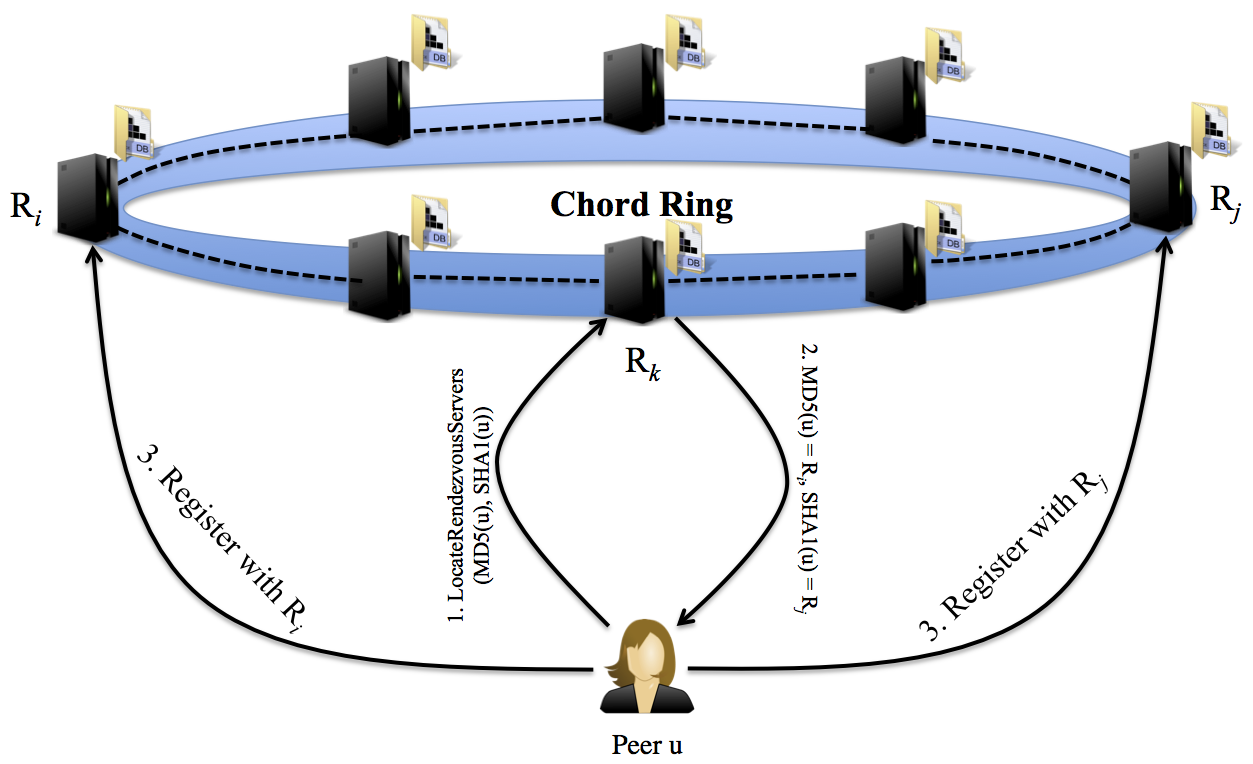}} \end{center}
\caption{Peer registration with rendezvous servers located on a chord ring. \label{dht_registration} }
\end{figure*}

We are assuming that a chord ring of rendezvous servers is already in place and a peer knows the IP address of at least one of the nodes on the ring. This can be done, perhaps through a publicly available website listing the IP addresses of some of these rendezvous servers. Figure \ref{dht_registration} shows the slightly modified registration process with the rendezvous servers.

Instead of registering directly with a rendezvous server, peer $u$ first, looks up the correct rendezvous servers on the DHT ring that correspond to its SHA-1 and MD5 hashed username values. Then, $u$ registers with those rendezvous servers using the same procedure described in Figure \ref{peer_registration_sd}. Note that as mentioned before, the only function that chord supports, is the look up function which returns the IP address of another node on the ring. Therefore, the added complexity in the global deployment model is very minimal in terms of the changes needed on the peer side.

\begin{figure} \begin{center} \scalebox{0.33}{\includegraphics{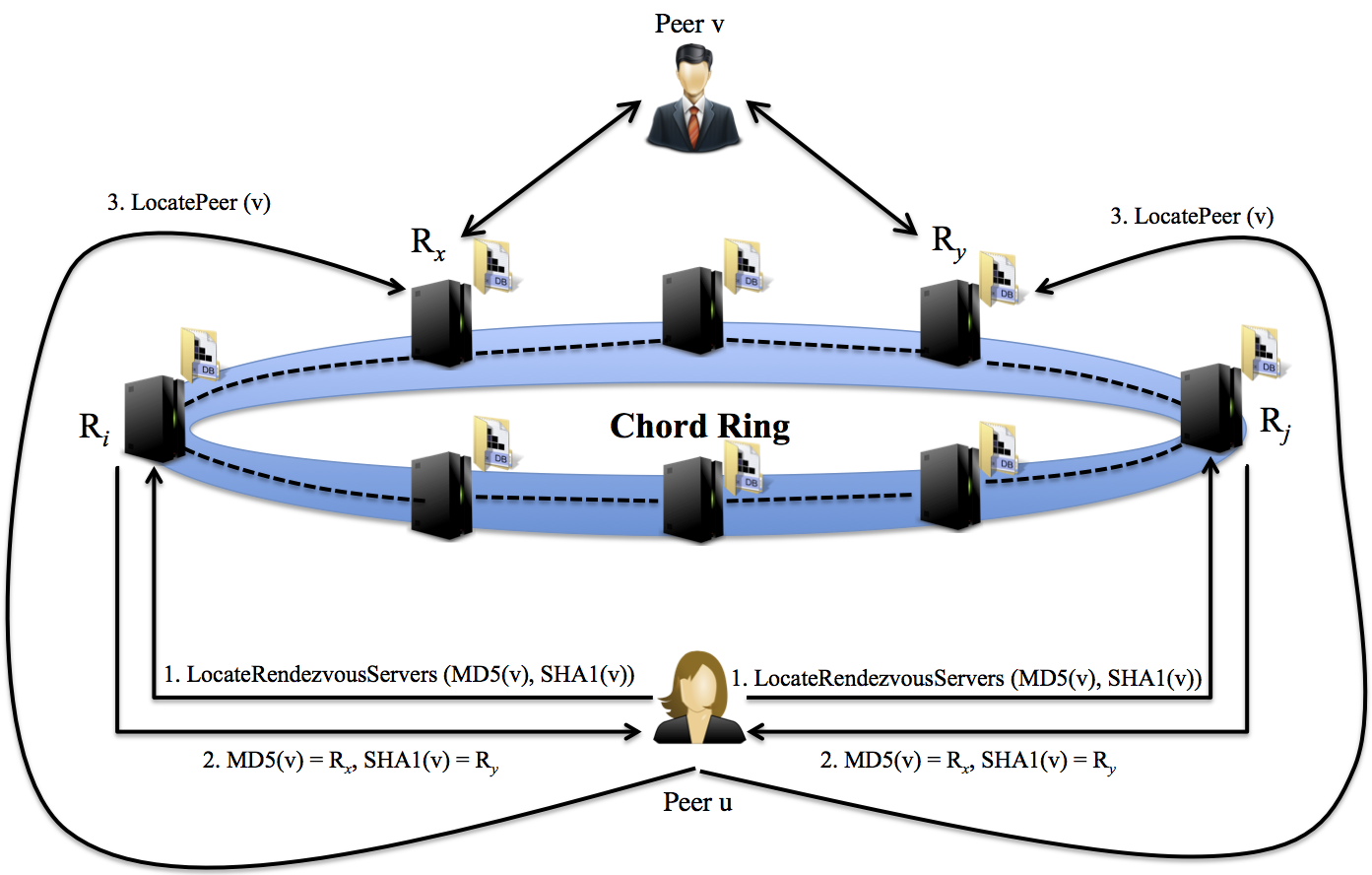}} \end{center}
\caption{Peer look up process on chord ring. \label{dht_lookup} }
\end{figure}

Finally, a peer can locate another peer by first, finding its corresponding rendezvous servers using a look up query on the chord ring and then, using the same process described in Figure \ref{locate_peer_sd} to obtain the connection information of the other peer as shown in Figure \ref{dht_lookup}.

\subsection{Application Layer}
\label{sec:applicationLayer}

 \begin{figure} \begin{center} \scalebox{1.1}{\includegraphics{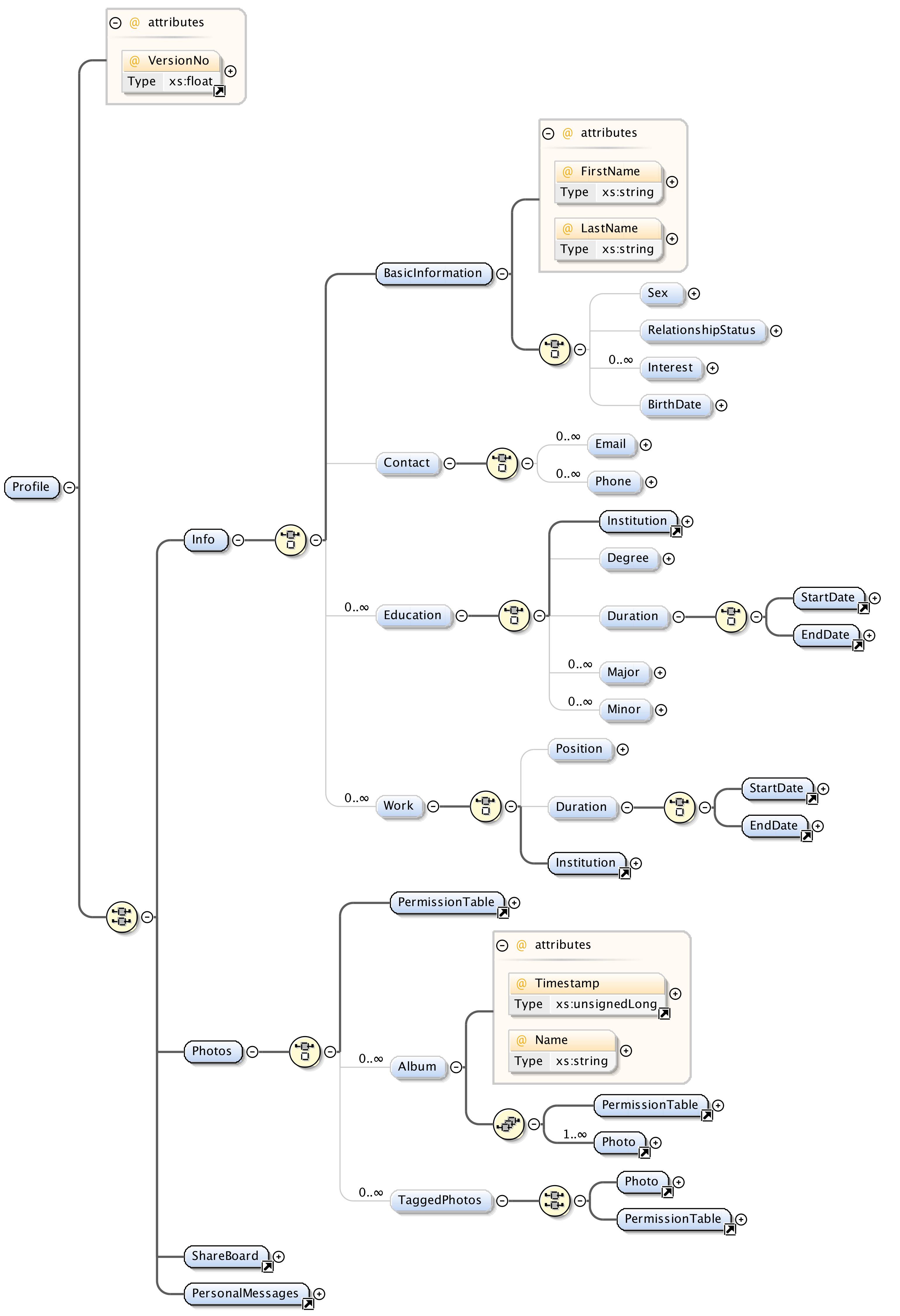}} \end{center}
\caption{XML schema for user profile. \label{profile_schema} }
\end{figure}

\begin{figure} \begin{center} \scalebox{1.5}{\includegraphics{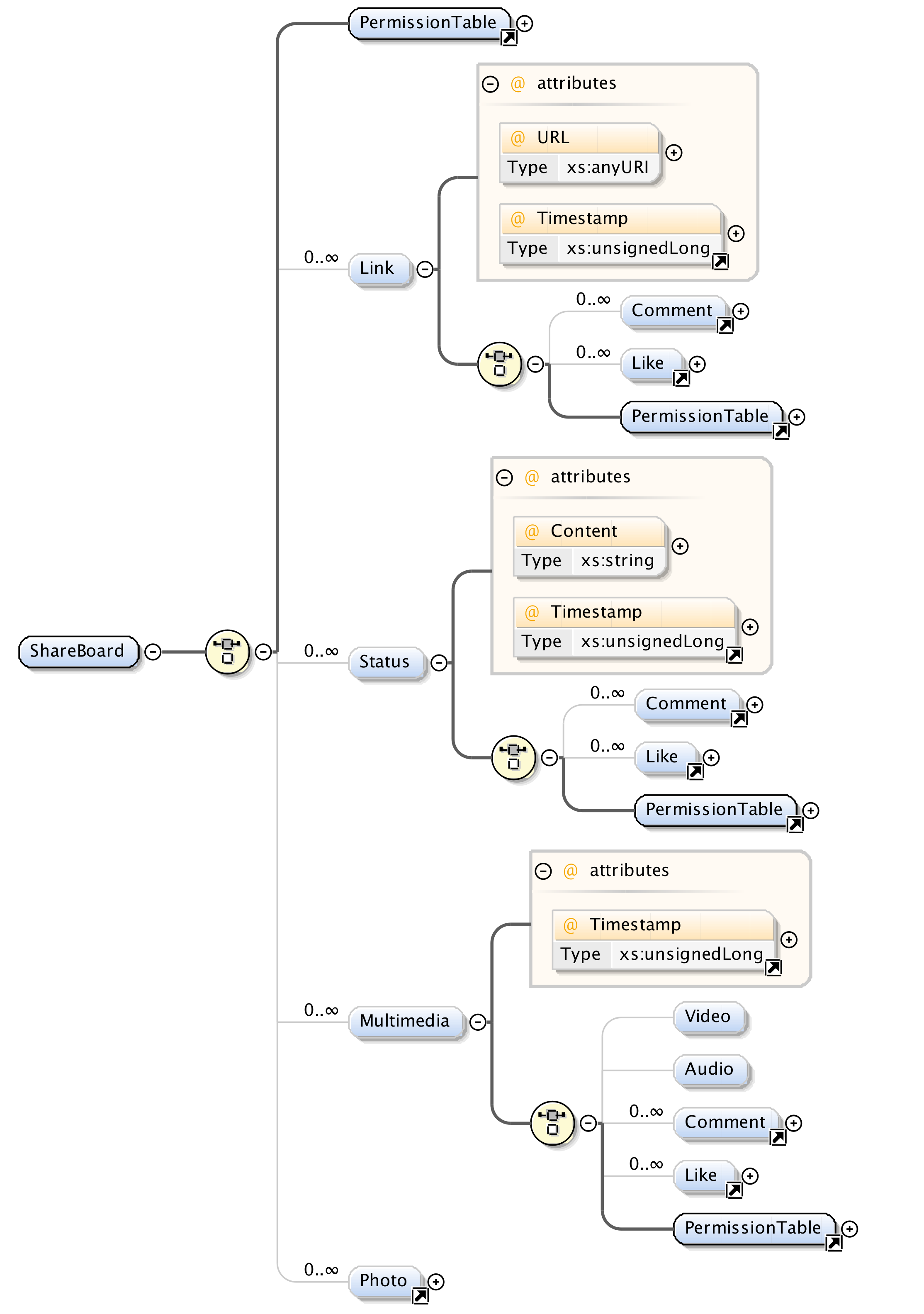}} \end{center}
\caption{XML schema for the share board component of the profile. \label{shareboard_schema} }
\end{figure} 

\begin{figure} \begin{center} \scalebox{1.1}{\includegraphics{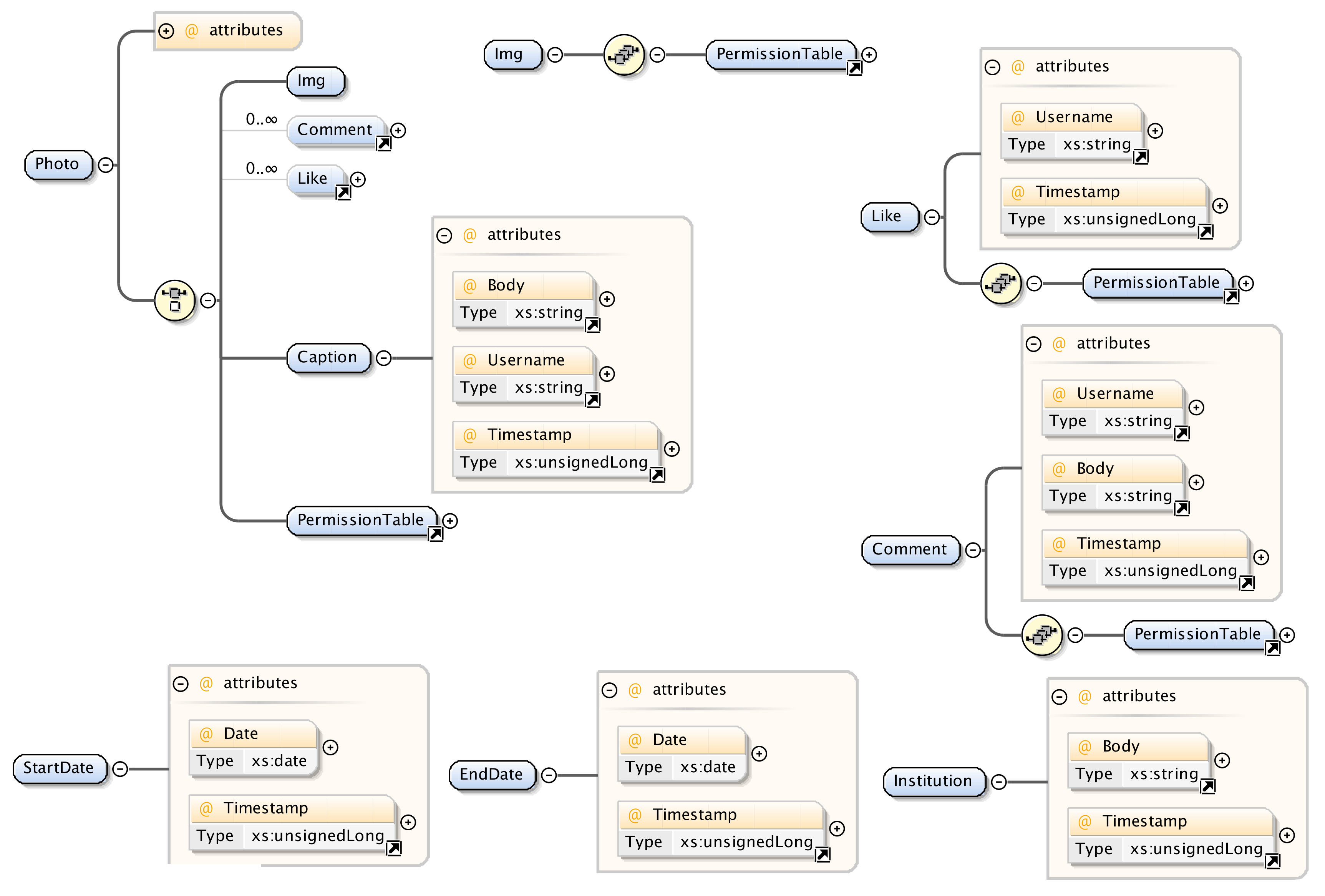}} \end{center}
\caption{XML schemas for other elements of user profile. \label{profile_components_schema} }
\end{figure}

\begin{figure} \begin{center} \scalebox{1.1}{\includegraphics{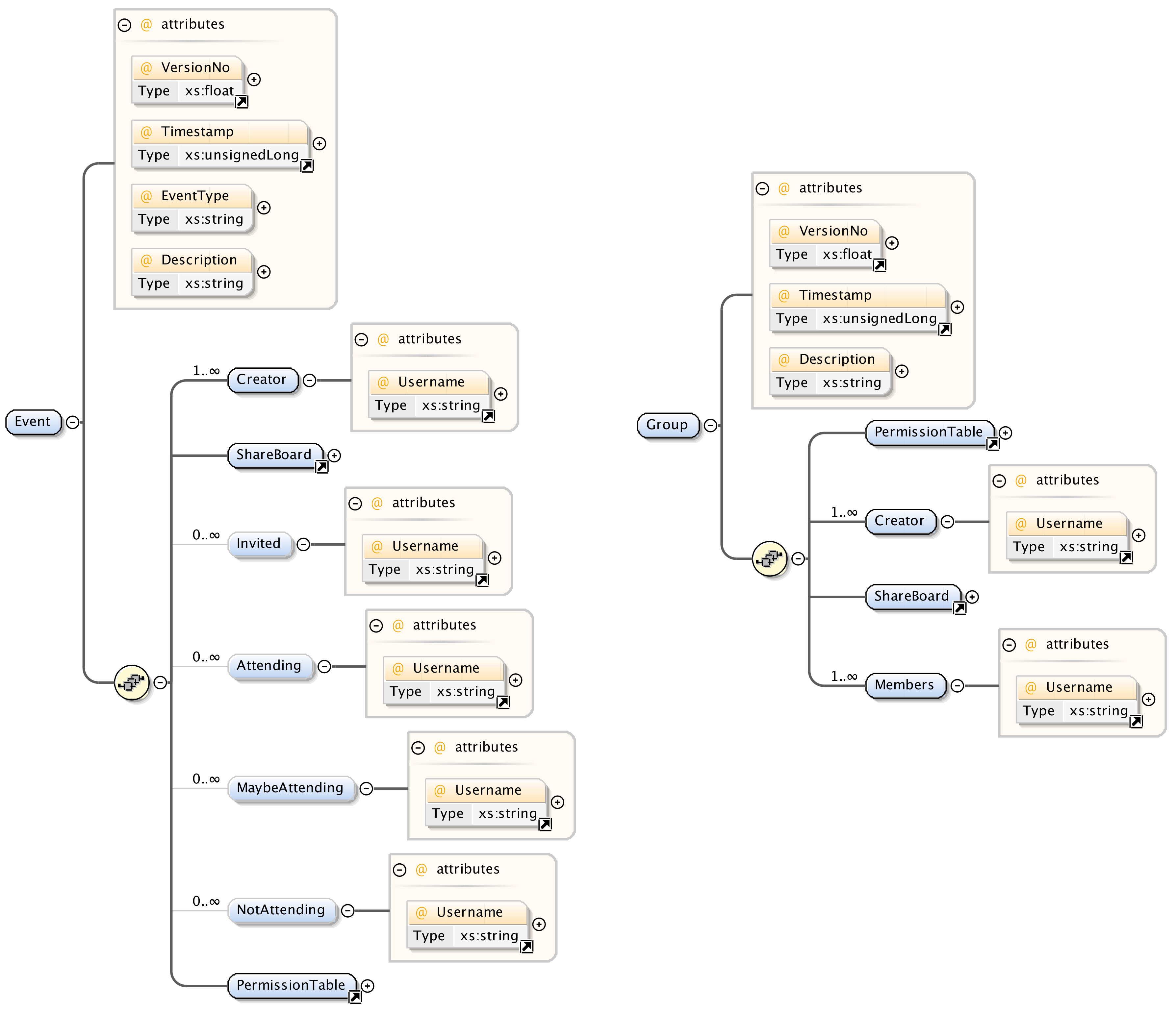}} \end{center}
\caption{XML schemas for event and group features. \label{event_group_schema} }
\end{figure}

\begin{figure} \begin{center} \scalebox{1.5}{\includegraphics{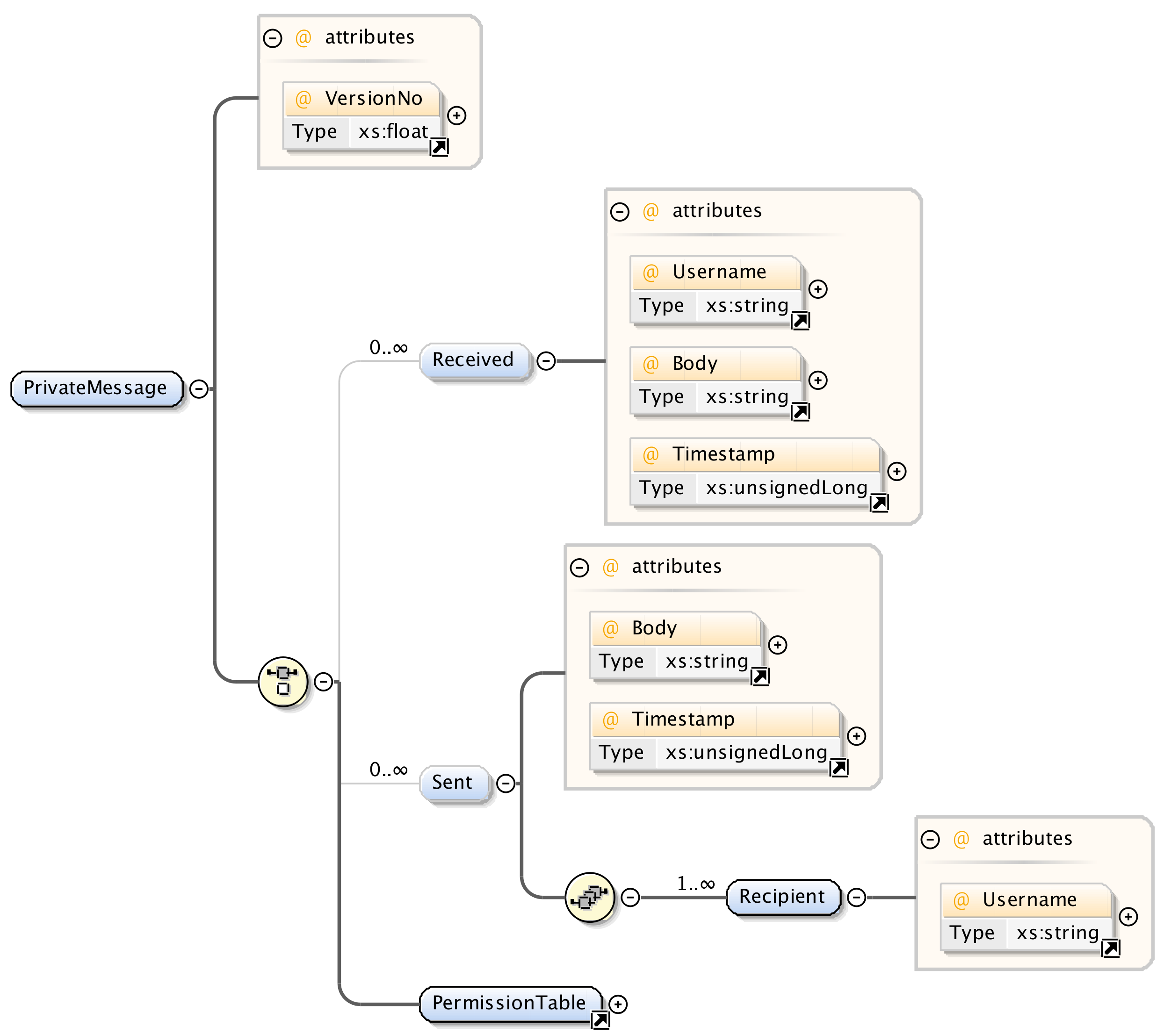}} \end{center}
\caption{XML schema for private message feature. \label{pm_schema} }
\end{figure}

\begin{figure} \begin{center} \scalebox{1.7}{\includegraphics{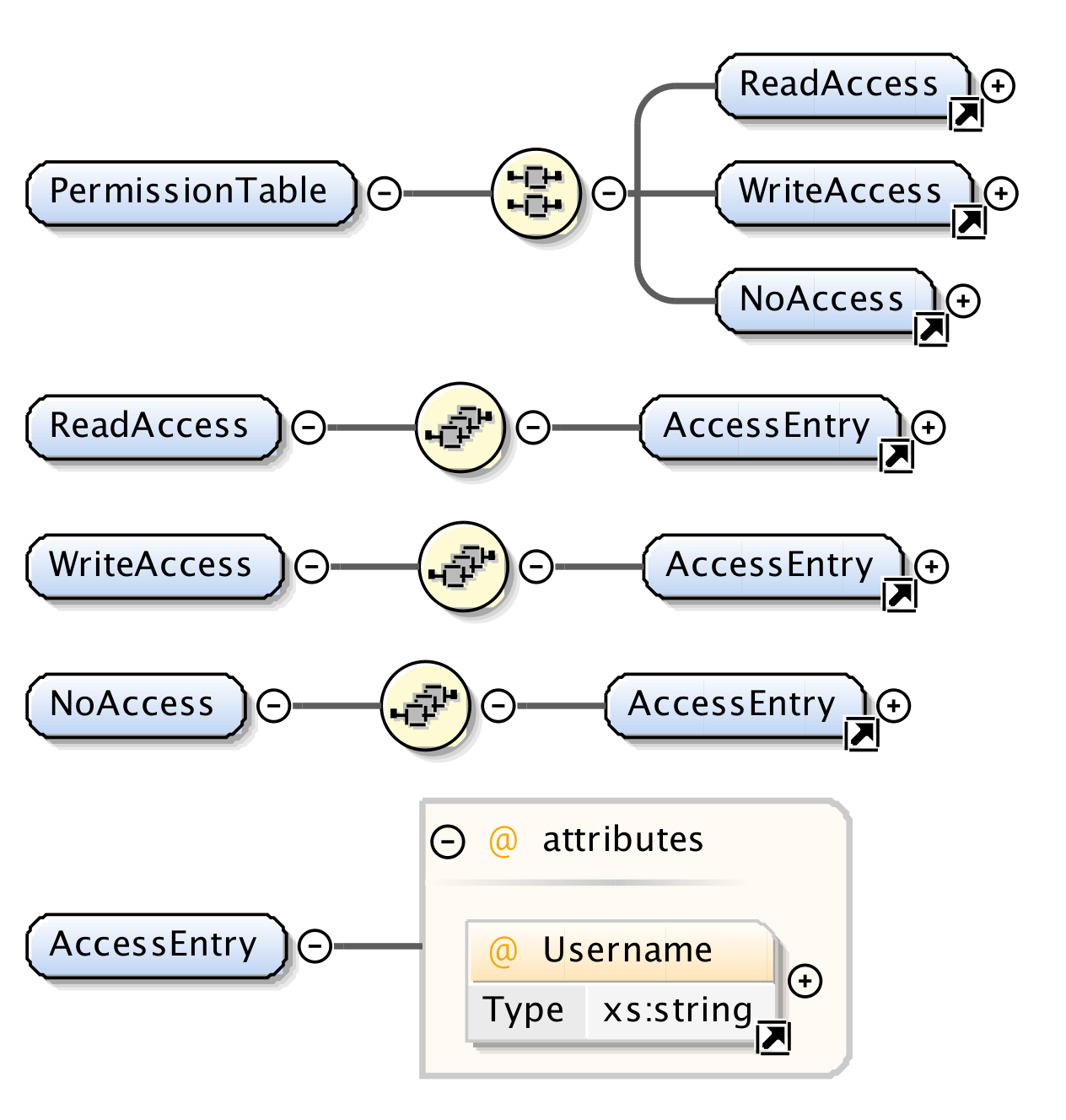}} \end{center}
\caption{XML schema for permission table. \label{permissionTable_schema} }
\end{figure}

\begin{figure} \begin{center} \scalebox{1.7}{\includegraphics{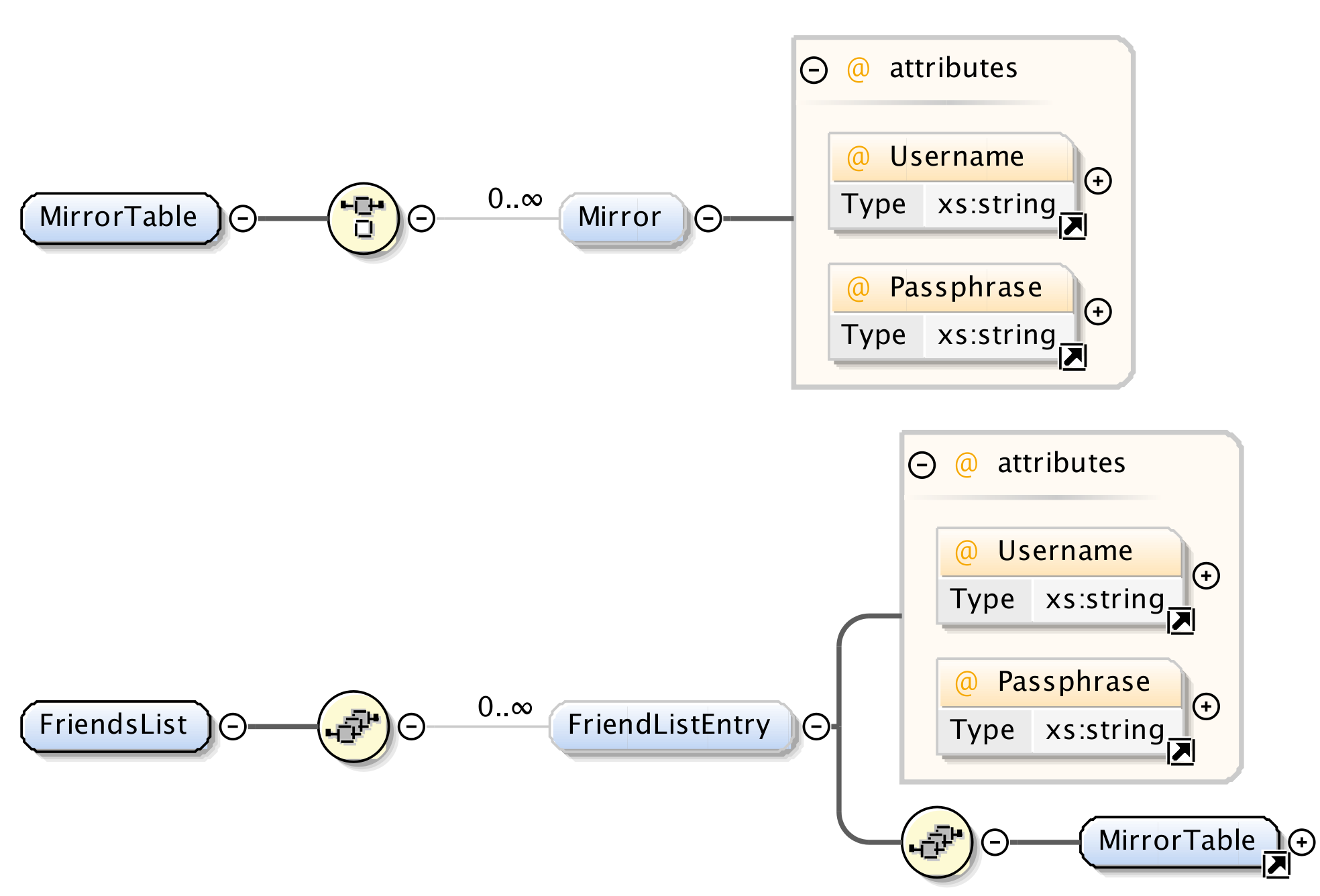}} \end{center}
\caption{XML schema for mirror and friend list tables. \label{mirror_friendlist_schemas} }
\end{figure}

This section describes the application layer of MyZone that is implemented on top of the service layer described in the last section. The application layer supports the following functionalities:
\begin{enumerate}
\item Implementing the user interfaces.
\item Enforcing permission policies defined by users, for different elements of their profiles. These policies define read, write, or no access permissions, for each element of the profile. 

The user may define these access permissions, per individual, or groups of users. This is very similar to the concept of access control used in operating systems. 
\item Replicating the user profiles on their mirrors to ensure that the most recent profiles are available to friends, even if the users are offline.
\item Finally, if a user agrees to serve as mirror for another user, the application layer needs to support the last two functionalities for the mirrored profiles.
\end{enumerate}

In addition to these, the application layer should be able to support a variety of features. This means that the application should easily adapt to changing features but in the mean time, the changes should not be forced onto the users. Hence, the application layer should be backward compatible.

We chose Extensible Markup Language (XML) to represent the profile data, permission and mirror tables and the friend list. XML is a set of rules for encoding documents in machine-readable form. It is widely used to represent arbitrary data structures. The simplicity, generality, and usability of XML, in addition to the wide range of programming languages that support it, has made it a perfect candidate to serve the purposes of the application layer design.

Figure \ref{profile_schema} shows the XML schema for a user profile. The blue ovals are elements of profile, and the orange boxes represent the attributes. The definitions of the complex components of profile are represented in the following Figures. 

Figure \ref{shareboard_schema} describes the XML schema for the share board component of the profile. The share board component is analogous to the Wall component of Facebook\cite{facebook} profiles. The remaining components of the user profile are described in Figure \ref{profile_components_schema}.

Figure \ref{event_group_schema} shows the XML schema for group and event features of the OSN, followed by the definition of the private message feature in Figure \ref{pm_schema}. Each element introduced in the previous schemas has a sub-element named permission table. 

The permission table defines the access permissions for that particular element. As you can see the permissions can be customized even at the lowest levels. The permission table has three parts: Read, Write, or No access. Each part includes the usernames with corresponding access permissions. The schema for the permission table is represented in Figure \ref{permissionTable_schema}.

Finally, the mirror and friend list tables are defined in Figure \ref{mirror_friendlist_schemas}. The schema for the mirror table includes the username and the corresponding passphrase of each mirror, while the friend list table includes the usernames and passphrases of friends and their corresponding mirror tables. When a friendship is revoked the username and passphrase of the corresponding user is removed from the friend list table. Also the mirror table is sorted based on the preference of the original user. When the user is offline her friends would try to access her mirrors based on this sorted mirror list. In addition to other friends the mirrors can be other devices belonging to the original users. For example the user may use her smartphone as the original profile storage, her laptop as the primary mirror and her desktop as the secondary mirror.

As described earlier in \ref{subsec:traffic_power}, traffic optimization and power management are one of the challenges facing the feasibility of peer to peer OSNs. This is mainly, because of the growing number of users that use their portable devices, e.g.  smartphones and tablet computers, to use the OSN services.

portable devices come with limited resources, i.e. traffic, bandwidth, and power, hence, the application should not be demanding of those resources. Since the data traffic contributes to all three, an essential requirement of the application layer is to use minimal traffic to reflect updates to friends. One logical approach towards this goal, is to use pulling, instead of pushing of information. This ensures that the user updates are only sent if there are any demands for them. 

To transmit minimal traffic, we use versioning to record changes to user profiles, events, groups, or private messages\footnote{The version number for the private message component is solely used for replication purposes, since the access to this component is available only to the profile owner and not even her mirrors. To enforce this access restriction, all private messages sent to mirrors, are encrypted using the user's public key.}. Each of these elements has its own version number attribute that is incremented upon every single change, while an {\it update log} stores the history of all updates and their corresponding version numbers on the host. 

When a user queries her friends for recent activities, she sends them the latest version numbers of their stored profile components. In return, she would receive only the missing updates, based on the update logs. This guarantees minimal traffic, which in combination with the pulling approach, results in minimal power consumption.

\section{Security Measures}
\label{ch:securityMeasures}

In this chapter, we explain the possible security attacks, the scenarios that implement them, and the measures taken to detect, prevent, and recover from these attacks. We start by identifying all the security attacks that apply to our peer to peer OSN environment and defining the scenarios that implement them. Then, we describe a modified peer registration process for the global deployment model, that embodies many of the security measures. Finally we explain how the design handles the attacks. 

The adversary model was defined earlier in \ref{subsubsec:adversary}. There, we described the malicious entities, their resources and the types of attacks that they can implement. Here, we further categorize the adversaries into three groups:
\begin{enumerate}
\item Those that try to steal user information i.e. profile, friend list, etc.
\item Those that try to isolate particular peers, or groups of peers from other users.
\item Those that try to make the system unresponsive by attacking its server components.
\end{enumerate}

As a reminder from earlier, we denote that a malicious node can be any component of the system with the exception of the certificate authority. This essentially means that any rendezvous or relay server or peer can be malicious\footnote{As mentioned earlier in \ref{subsubsec:trustmodel}, a peer would not act maliciously towards its friends}. Note that these assumptions only apply to the global deployment model, as opposed to the local deployment model where the goal of constructing a private OSN would require the components to be trustworthy. 

In addition, any malicious node can choose selectively to act correctly or maliciously towards other components. Based on these assumptions any attack that attempts to use impersonation e.g. man in the middle, spoofing, eavesdropping etc. will not be a concern since we assume that all components of the system except the certificate authority and friends are not trusted. However, the design uses a decentralized technique that detects and isolates the malicious entities over time, as it will be explained later on.

\subsection{Security Attacks and Scenarios}
\label{subsec:scenarios}

The security attacks that apply to our model and their corresponding scenarios are:
\begin{itemize}
\item Sybil attacks: an attacker creates a large number of identities and dominates the overlay network by fooling the protocols, and subverting mechanisms based on redundancy.

Scenarios:
\begin{itemize} 
\item S.1) A malicious node creates virtual rendezvous servers and tries to take over a particular key (username) or pollute other rendezvous server tables.
\end{itemize}
\item Eclipse attacks also known as routing table poisoning: an attacker controls a sufficient fraction of the neighbors of a node, hence the node can be ÒeclipsedÓ. This kind of attack applies to network proximity based DHTs like Pastry\cite{pastry} and Tapestry\cite{tapestry} and is very hard to achieve in Chord. 

Scenario:
\begin{itemize}
\item E.1) A set of malicious rendezvous servers try to target (eclipse) a particular correct rendezvous server by poisoning the routing tables of other rendezvous servers.
\end{itemize}
\item Routing attacks: an attacker may do a combination of the following:
\begin{itemize}
\item Refuse to forward a lookup request. 
\item It may forward it to an incorrect, non-existing, or malicious node. 
\item It may pretend to be the node responsible for the key.
\end{itemize}

Scenarios:
\begin{itemize}
\item R.1, R.2, R.3) Malicious rendezvous servers do one or some combination of the above, on a locate rendezvous server request.
\end{itemize}
\item Storage attacks: a node routes requests correctly, but denies the existence of a valid key or provides invalid data as response. This attack can be in conjunction with the routing table attack.

Scenarios:
\begin{itemize}
\item T.1) A rendezvous server returns incorrect information in response to a locate peer request.
\end{itemize}
\end{itemize}
 
\subsection{Security Measures}
 \label{subsec:measures}
 
\begin{figure} \begin{center} \scalebox{0.7}{\includegraphics{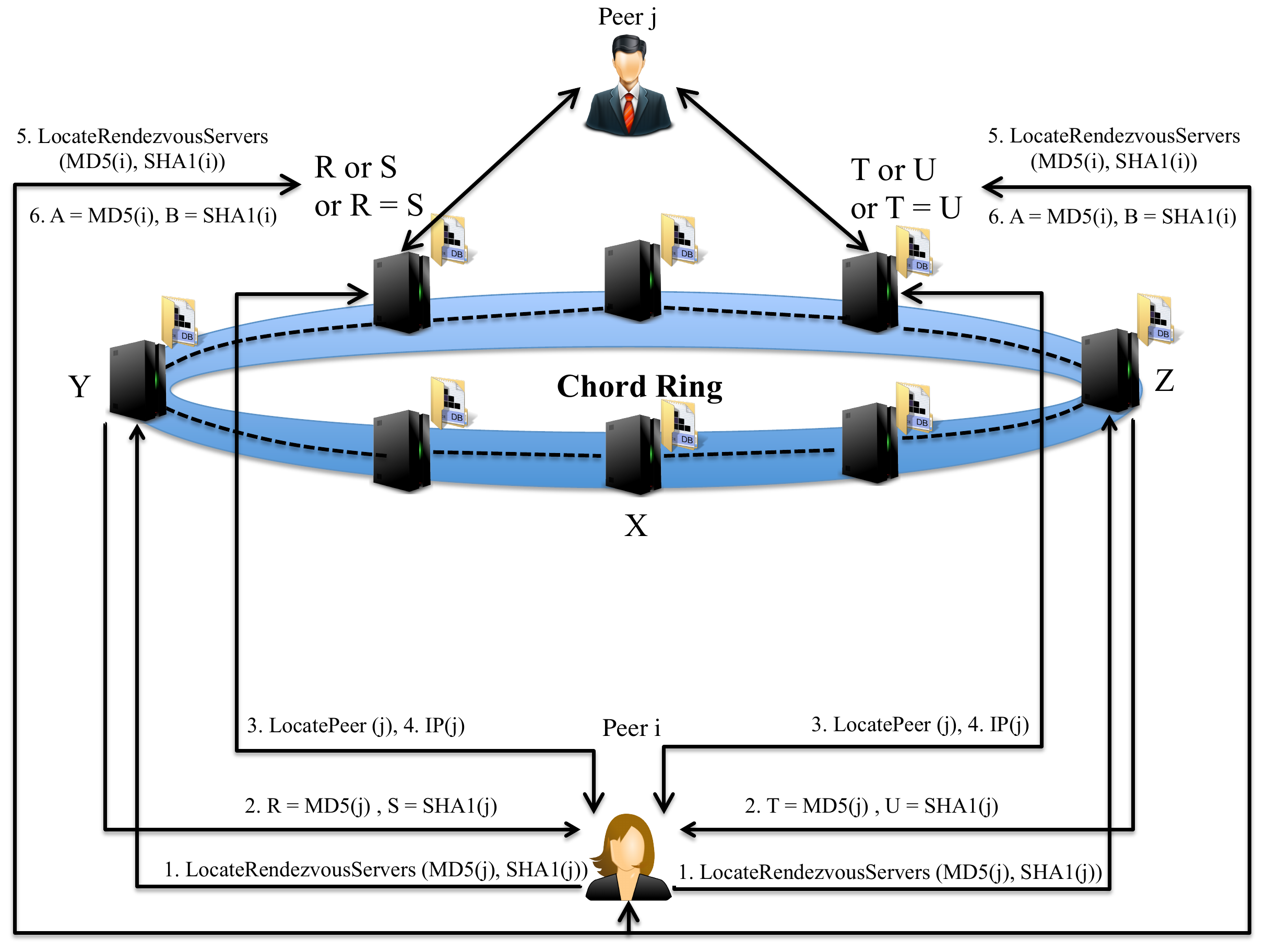}} \end{center}
\caption{The modified peer registration process for the global deployment model. \label{modified_dht_registration} }
\end{figure}

Now that we have defined the security attacks and their corresponding scenarios, we can introduce the modified peer registration process. This modification only applies to the global deployment model that uses chord DHT. Figure \ref{modified_dht_registration} shows the overall registration process in the global deployment model.

The process is modified to reflect the fact that although the peers initially assume that the rendezvous servers are acting correctly, they don't trust them until their behaviors are verified by another correct entity. Rendezvous server $X$ corresponds to the initial rendezvous server $R_k$ in Figure \ref{dht_registration}, while $Y$ and $Z$ correspond to the rendezvous servers {\it assumed to be} responsible for MD5 and SHA-1 hashed username values respectively. 

The modified process is executed before peer $i$ registers with the rendezvous servers. Using this process peer $i$ can  verify if $Y$ and $Z$ are really responsible for its keys, or if they are malicious nodes trying to take over $i$. The process can be summarized as the following steps:

\begin{enumerate}
\item $(R, S) = Y.locateRendezvousServers(${\it MD5(j)}$, ${\it SHA-1(j)}$)$; // Assuming that $j$ is a friend of $i$: $R$ and $S$ are rendezvous servers responsible for MD5 and SHA-1 username keys respectively. 

$(T, U) = Z.locateRendezvousServers(${\it MD5(j)}$, ${\it SHA-1(j)}$)$; // Shown as steps 1 and 2 in Figure \ref{modified_dht_registration}.
\item $IP_j(R) = R.locatePeer(j.passphrase)$; $IP_j(S) = S.locatePeer(j.passphrase)$; // $IP_j(R)$ represents the IP address of $j$ returned by rendezvous server $R$.
	 
$IP_j(T) = T.locatePeer(j.passphrase)$; $IP_j(U) = U.locatePeer(j.passphrase)$; // Shown as steps 3 and 4 in Figure \ref{modified_dht_registration}. 
\item if ($i.connectTo(IP_j(R)) == false$ \&\& $i.connectTo(IP_j(S)) == false$ 

 \&\& $i.connectTo(IP_j(T)) == false$ \&\& $i.connectTo(IP_j(U)) == false$) then \{// none of the returned IPs for $j$ is correct.
\begin{enumerate}
\item find another rendezvous server.
\item goto $1$.

\}else\{
\item $(A, B) = D.locateRendezvousServers(${\it MD5(j)}$, ${\it SHA-1(j)}$)$; // Steps 5 and 6. $D$ is any of $\{R, S, T, U\}$ that returned the correct IP address for $j$ and is considered as a correct rendezvous server.
\item If ( $A == Y$ ) then $Y$ is the correct rendezvous server and $X$ acted correctly;
	 	 
\item else $Y$ is not the correct rendezvous server and $X$ and $Y$ are malicious nodes.

goto $1$.
	 	 
\item If ( $B == Z$ ) then $Z$ is the correct rendezvous server and $X$ acted correctly;
	 	 
\item else $Z$ is not the correct rendezvous server and $X$ and $Z$ are malicious nodes.

goto $1$.

\}
\end{enumerate}
\end{enumerate}

The above process can only be used after user $i$ is friend with at least one other user. Next, we explain how the overall design addresses different attack scenarios by prevention, detection, and recovery techniques. 

\subsubsection{Prevention}

Out of attack scenarios described in \ref{subsec:scenarios}, there are three scenarios that are prevented by our design:

\begin{itemize} 
\item A malicious node targets a particular peer to either render the user unreachable by DDoS attack or hijack information from the user. 

Solution: This requires the attacker having access to the user connection information. The connection information can be obtained in two ways: sending the correct passphrase of the user to the rendezvous server responsible for the target user, or through registering with a malicious rendezvous server. 

The passphrases are only shared with friends that are all trusted and will not act maliciously. Furthermore, targeting a specific peer by a malicious rendezvous server is very unlikely due to the key distribution properties of consistent hashing\cite{consistent_hashing}. 

Even if a malicious node somehow obtains the connection information of the peer, although a DDoS attack is feasible, unauthorized access to the user information is prevented. This is because, the connection is encrypted using symmetric encryption and the procedure used to exchange the session key uses keys from both parties. 

The key exchange procedure ensures that the session key is encrypted by the sending party, in a manner, that it can only be decrypted by the receiving party. This, guarantees authentication and integrity at the same time, preventing man in the middle and impersonation attacks. In addition, connections from non-friend peers are refused.

\item S.1) A malicious node creates virtual rendezvous servers and tries to take over a specific user or pollute the rendezvous server tables to target that user.

Solution: the consistent hashing guarantees the balanced distribution of the identifiers. Consequentially, although an attacker can create a lot of malicious virtual rendezvous servers, it can only take over a specific user $x$, if it can predict the IP address of $Successor(x)$. This requires the malicious user to be able to reverse the consistent hashing function which is known to be hard. 

\item E.1) A set of malicious rendezvous servers try to bypass (eclipse) a particular correct rendezvous server by poisoning the routing tables of other rendezvous servers.

Solution: Although this attack is not explicitly prevented, it is effective only in DHTs that are based on network proximity and does not apply to chord\cite{dht_sec_survey}. Furthermore, the two independent identifiers for each username i.e. MD5 and SHA-1, would very likely result in two independent paths to two different rendezvous servers. This makes eclipse attacks even less effective. 

\end{itemize}

\subsubsection{Detection}

Our design would be able to successfully detect the following attack scenarios as explained below:

\begin{itemize}
\item R.1) Malicious rendezvous server refuses to forward a locate rendezvous server request.

Solution: The peer tries to find another rendezvous server from a list of publicly available rendezvous servers until it receives a reply.
\item R.2, R.3) Initial malicious rendezvous server $X$ returns malicious nodes $Y$ and $Z$ in response to $LocateRendezvousServers(${\it MD5(i)}$, ${\it SHA-1(i)}$)$ sent by peer $i$ in the initial stage of registration.

Solution: Malicious behavior at the registration stage is detected using the process described in \ref{subsec:measures}. 

\item T.1) A rendezvous server returns incorrect information in response to a locate peer request.

Solution: Incorrect registration information can be identified by comparing the hash value computed over $IP + Port + \dots$ with the signed hash value i.e. $E(Q_{peer})[IP + Port + \dots]$ introduced in section \ref{sec:serviceLayer}.

This mechanism can also be used by rendezvous servers to detect malicious rendezvous servers upon replication of data. This is done by each rendezvous server, verifying the authenticity and integrity of the registration information that it receives from other servers, before replicating it. If a rendezvous server receives a modified registration information, it will not forward or replicate it.

\end{itemize}

\subsubsection{Recovery}

The biggest part of recovery is to isolate the malicious rendezvous servers after detection. The malicious rendezvous servers are removed by a collaborative decision making that uses a reputation based scheme. There are two ways to detect a malicious node: At the time of peer registration as described in section \ref{subsec:measures}, and After registration. 

Peer $i$ can detect malicious behavior of rendezvous server $X$ that it registered with, by receiving notifications from its friends who received incorrect or out of date registration information  from $X$, in response to {\tt locatePeer($i$)} queries. Note that in this scenario, $X$ has behaved correctly during the initial registration process otherwise $i$ would not have registered with $X$ from the beginning. 

Peer $i$ would send a complaint to a correct rendezvous server found in steps $3$ and $4$ in the registration process described in section \ref{subsec:measures}. The complaint is spread on the chord ring and is recorded by the nodes that have an entry for $X$ in their routing tables. 

Server $X$ would be removed from the chord ring, only if the ring receives enough distinct complaints specifically from peers that are registered with $X$. The distinct complaints are required in order to prevent malicious peers from sending out falsified complaints, hence, removing a correct rendezvous server. 

If the received complaints regarding $X$ are not enough, the servers on the chord ring would agree not to isolate $X$, as it will be considered more beneficial than harmful. The benefit is the result of selective malicious behavior of $X$ towards peers. Respectively, peer $i$ would send a complaint only if it receives enough number of notifications from its friends. If these notifications are few, peer $i$ deducts that the benefit of $X$ still outdoes its harm. 

The complaints sent by any peer need to be authenticated. This means that the entire complaint needs to be signed by the peer and it should include a timestamp to prevent replay complaints. 

The described scheme is designed based on the following goals:

\begin{enumerate}
\item A malicious rendezvous server can't complain about a correct rendezvous server and only a peer registered with the rendezvous server can complain about it.
\item A malicious peer can only complain about its own rendezvous server and only once. Therefore, if $r$ is the minimum number of complaints that are needed to exclude a node from the ring, then $r$ of these malicious peers need to register with the same rendezvous server to remove a specific rendezvous server. 
\item A peer can falsely believe that her rendezvous servers are malicious, only if her friends intentionally send notifications to her. This would result in the unreachability of the peer and would violate the assumption that a peer would not act as an adversary to her own friends.
\end{enumerate}

\section{Implementation}
\label{ch:implementation}

\begin{figure} \begin{center} \scalebox{.3}{\includegraphics{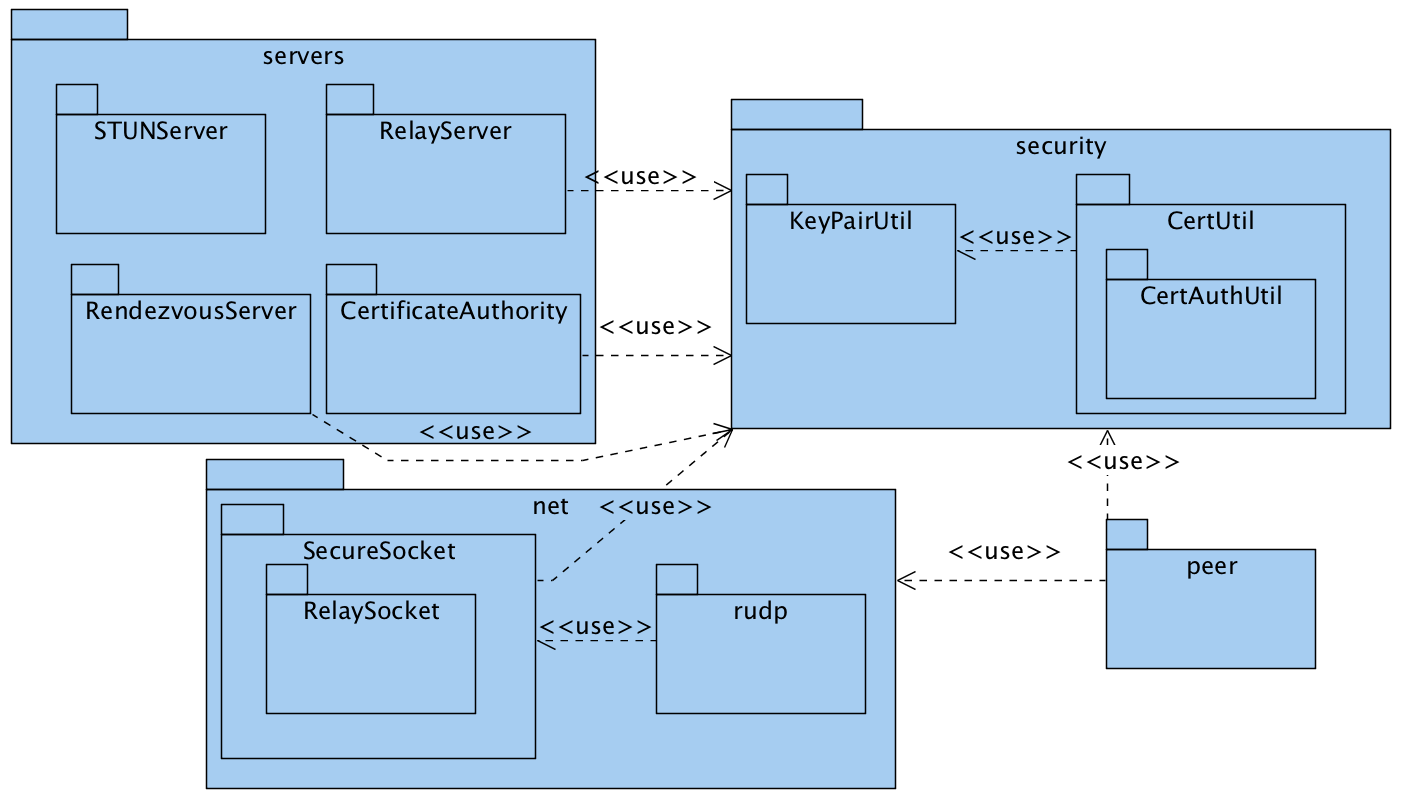}} \end{center}
\caption{Overall view of all the composing packages of the service layer and their relationships. \label{overall_package} }
\end{figure}

In this chapter, we explain the structure of our service layer code and define the interfaces provided by each component. As described earlier in section \ref{sec:serviceLayer}, the service layer includes five types of components, namely, Certificate Authority, Relay, Rendezvous and STUN Servers and Peer. The corresponding Java code for the service layer, is structured in a way that reflects the distinctions of these entities. As shown in Figure \ref{overall_package}, the overall view of the service layer code lays out four different packages: \begin{itemize}
\item servers: Which includes the implementation of the four server components of the service layer, each in a separate package.
\item net: Which implements a secure socket over both UDP and TCP. This package includes SecureSocket and rudp packages. rudp package implements reliable UDP. The SecureSocket package implements a secure socket over TCP and reliable UDP, while its sub package RelaySocket, implements a relayed secure socket.  
\item security: Which implements security utilities needed by the other components.
\item peer: Which implements the services available to peers.
\end{itemize} 

\begin{figure} \begin{center} \scalebox{.3}{\includegraphics{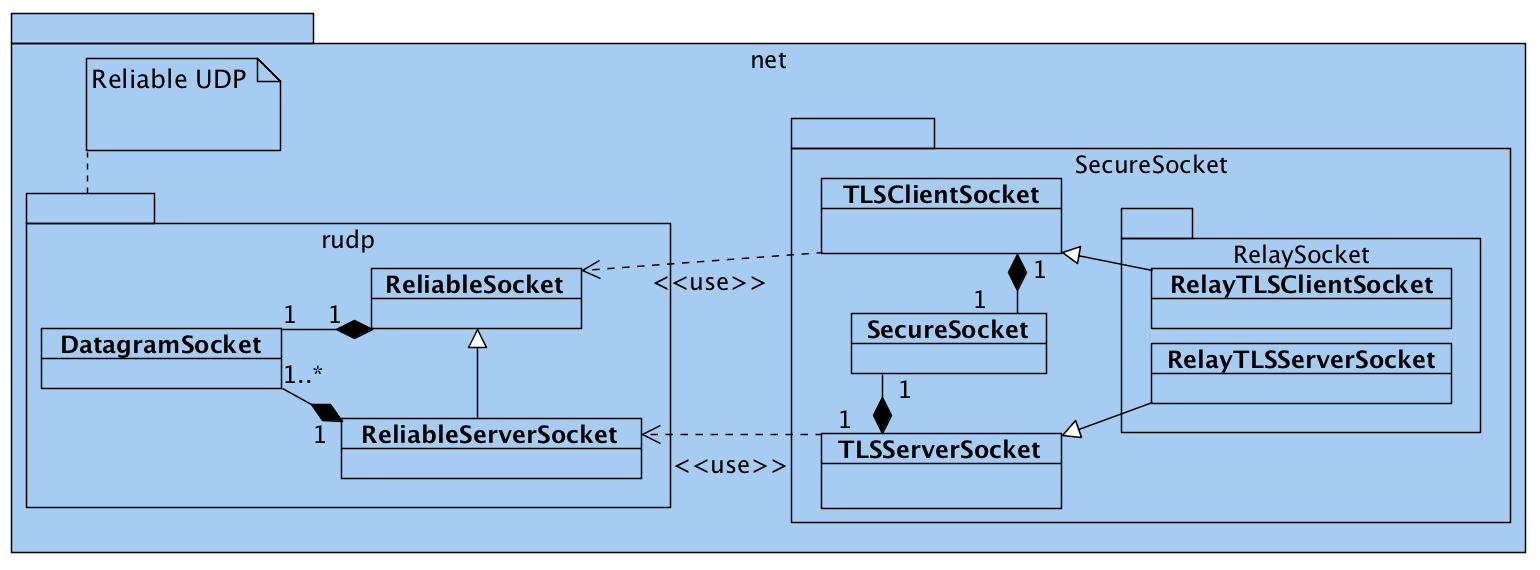}} \end{center}
\caption{Class diagram for the net package. \label{net_package} }
\end{figure}

\begin{figure} \begin{center} \scalebox{.3}{\includegraphics{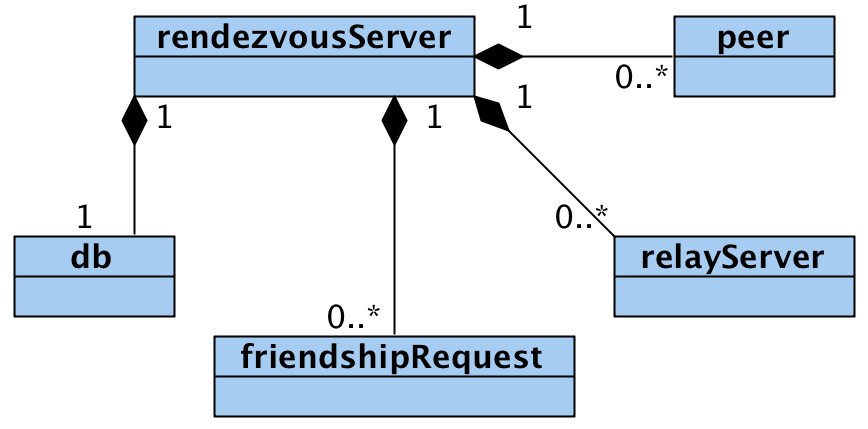}} \end{center}
\caption{Class diagram for the RendezvousServer package. \label{rendezvous_package} }
\end{figure}

\begin{figure} \begin{center} \scalebox{.3}{\includegraphics{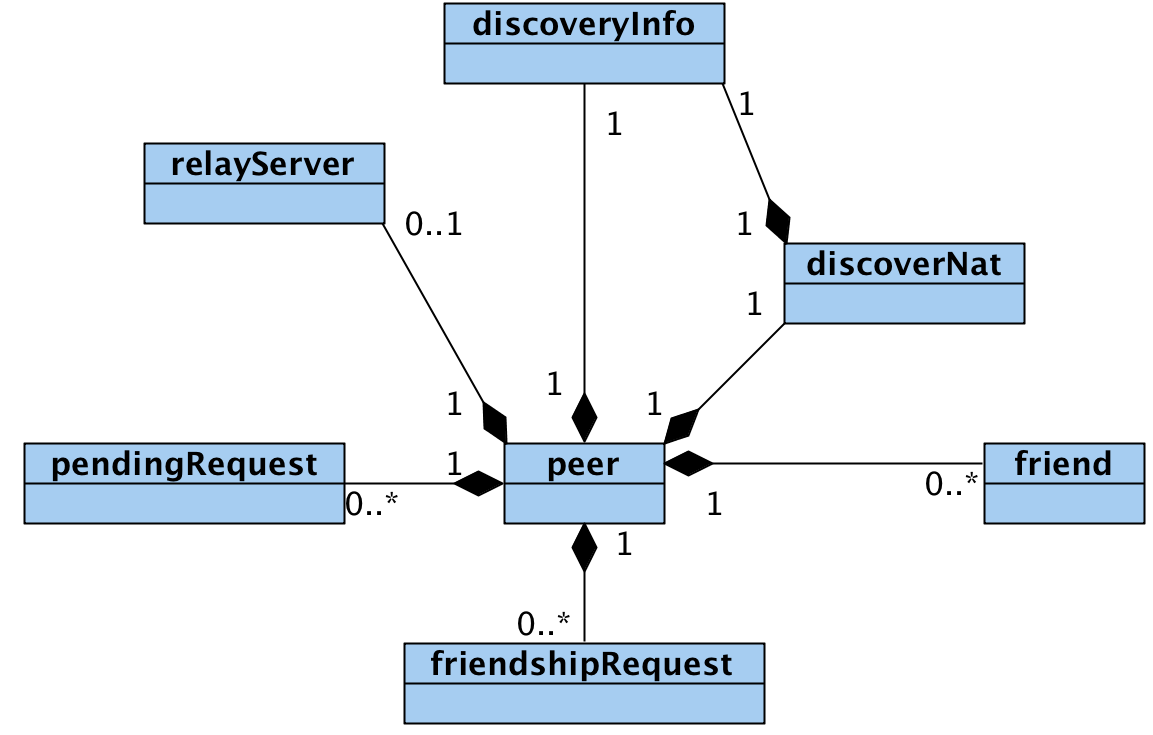}} \end{center}
\caption{Class diagram for the peer package. \label{peer_package} }
\end{figure}

Figure \ref{net_package}, \ref{rendezvous_package} and \ref{peer_package} describe net, RendezvousServer and peer packages in more detail using UML class diagrams. 
The net package includes the rudp and SecureSocket packages. The rudp package is implemented using two classes: The ReliableSocket class implements a reliable datagram client socket using the standard datagram socket provided by Java. The ReliableServerSocket class implements a reliable datagram server socket using the ReliableSocket class. 

The SecureSocket package is composed of TLSClientSocket, TLSServerSocket, and SecureSocket classes and RelaySocket package. The TLSClientSocket and TLSServerSocket implement client and server side  {\it TLS like} secure sockets respectively. The main functionalities of these two classes is to perform the certificate and session key exchanges required as part of TLS handshake. 

Upon successful exchange of the session key, a SecureSocket class is instantiated by each of these classes. The SecureSocket class performs send and receive functionalities, while encrypting all transmitted data using the exchanged session key. The RelaySocket package implements the same functionalities as TLSClientSocket and TLSServerSocket but specifically for a relayed connection. As mentioned earlier, the relayed connection is an end to end secure connection and can not be decrypted by the relay server.

The RendezvousServer package includes rendezvousServer class that implements the functionalities of the rendezvous server. The db class implements the database driver needed by the RendezvousServer class to communicate with the rendezvous server database. The other classes are used to store data for peer, relay server and friendship requests. 

Finally, the peer package implements all the functionalities available to peers using the peer class. The peer class uses discoverNat to discover the type of NAT that the peer is behind, and stores this information in the discoveryInfo class. As for the RendezvousServer package, all the other classes are used to represent optional data.

Tables \ref{interfaces_p1}, and \ref{interfaces_p2} define all the interfaces for each of the components. The service layer code is the main part of our implementation which is written in Java using {\it only} the standard functions available, and is composed of over seventeen thousand lines of code.

\begin{sidewaystable}
\caption{Description of the interfaces for Rendezvous Server, and Relay Server components of the service layer.} \label{interfaces_p1}
\centering
\begin{tabular}{|p{3.5cm}|p{5cm}|p{13cm}|}
\hline
Component & Interface & Description \\
\hline Rendezvous Server & {\tt RendezvousServer(int port, String dbUsername, String dbPassword, String dbURL, String sessionKeyType, String asymAlgorithm, String sessionCipher Algorithm, int age, int refreshInterval)} & 

$-$ port: the port number that this rendezvous server listens on.

$-$ dbUsername: the username for the mysql\footnote{We use mysql dbms for rendezvous server database.} account.

$-$ dbPassword: the password of the corresponding username for the mysql account.

$-$ dbURL: the url to the mysql database.

$-$ sessionKeyType: the type of the symmetric encryption key used for session encryption

$-$ asymAlgorithm: the type of the algorithm used for asymmetric encryption.

$-$ sessionCipherAlgorithm: the type of the symmetric encryption algorithm used for session encryption.

$-$ age: the maximum period of time that a relay server is kept in the relay servers table without being refreshed.

$-$ refreshInterval: the refresh/update interval for relay servers. This value needs to be less than the value of age.
\\
\hline Relay Server & {\tt RelayServer(String rendezvousSrvAddr, int rendezvousSrvPort, int port, int maxConnections, int pingInterval)} &

$-$ rendezvousSrvAddr: the address of the rendezvous server which this relay would register with. 

$-$ rendezvousSrvPort: the remote port to connect on the rendezvous server.

$-$ port: the port that the relay server would be listening on. 

$-$ maxConnection: the maximum number of serving connections that this relay can handle.

$-$ pingInterval: at the end of this interval each serving socket needs to send a {\tt SERVER\_IS\_ALIVE} packet so that the relay server keeps it registered in its connection list.
\\
\hline
\end{tabular}
\end{sidewaystable}

\begin{sidewaystable}
\caption{Description of the interfaces for Certificate Authority Server, STUN Server, and Peer components of the service layer.} \label{interfaces_p2}
\centering
\begin{tabular}{|p{3.5cm}|p{6cm}|p{11cm}|}
\hline
Component & Interface & Description \\
\hline Certificate Authority Server & {\tt CAServer(int port, int backlog, InetAddress bindAddr, String caName, String priKeyPath, String keyAlgorithm)} &

$-$ port: the port number that the server listens on.

$-$ backlog: the number of waiting clients in the backlog queue.

$-$ bindAddr: the InetAddress of the network interface that the server is going to use.

$-$ caName: the name of the certificate authority. This is mainly used by the CA to read its private and public keys.

$-$ priKeyPath: the path to the directory where the private key of CA is stored at.

$-$ keyAlgorithm: the algorithm that is going to be used to issue certificates.
\\
\hline STUN Server & {\tt StunServer(int primaryPort, InetAddress primary, int secondaryPort, InetAddress secondary)} & 

$-$ primaryPort: The port number for the first network interface.

$-$ primary: the inet address of the first network interface.

$-$ secondaryPort: The port number for the second network interface.

$-$ secondary: the inet address of the second network interface.
\\
\hline Peer & {\tt peer(String username, int port, String myCertPath, String caCertPath, String myKeyPath, String passphrase, byte[] mirrors)} & 

$-$ username: The username of the user.

$-$ port: The port number that the peer accepts connections on. 

$-$ myCertPath: The path to the directory that holds the user certificate.

$-$ myKeyPath: The path to the directory that contains the key pair of the user.

$-$ passphrase: The passphrase of the user. 

$-$ mirrors: The encrypted mirror table of the user. \\
\hline
\end{tabular}
\end{sidewaystable}

\section{Conclusion}
\label{ch:conclusion}

The busy lifestyle of growing number of people has made conventional ways of socializing a luxury not available to all. Online social networks have made it possible for these people, to extend their social connections, while maintaining the existing ones. This has made OSNs a huge success with hundreds of millions of people using their services on daily basis. The large number of users has transformed OSNs into a more effective social media to spread ideas, news, opinions and more, comparing to conventional ones. 

Despite all their benefits, OSN service providers have been known to violate user privacy by selling user information to market researchers and other businesses. This has raised a lot of concerns in recent years, as people are becoming more and more aware of it and frequent changes in user privacy policies has only contributed to this. 

Furthermore, their growing use as a powerful social media has made them an effective tool to organize popular movements. This has made them a target of different attacks by the opposing entities. These attacks vary from traffic filtering to denial of service attacks to hijacking user information. 

Finally, ever-changing user interfaces and growing number of service providers have only made the user experience more frustrating. These shortcomings have motivated us to propose a peer to peer design for a next-generation OSN called MyZone. To our knowledge, MyZone is the first OSN with the following properties:
\begin{itemize}
\item It preserves user privacy by storing user information on their own devices and replicating them on a number of other devices belonging to trusted friends. 
\item It is secure based on a {\it ``need to know basis''} philosophy and all connections are encrypted.
\item It is resilient to wide spread attacks and outages performed by entities as powerful as governments.
\item The local deployment model can be set up conveniently and quickly to construct a private OSN for a limited number of users.  
\item It is backward compatible and a wide variety of client applications with different user interfaces and features can coexist while users can choose the applications that fit their needs and preferences. 
\end{itemize}

We described our two layered design in details and introduced security measures that makes it resilient under different attack scenarios. This was followed by describing the structure of the service layer code and defining the interfaces for each of these services. At this stage, we are designing the user interfaces, and implementing the client application, which will be available in the near future.

The availability of user profiles even when the users are offline, was identified as perhaps the most serious obstacle faced by any peer to peer OSN. Our design addressed this by selecting mirrors among friends. Selecting the most appropriate friends as mirrors, is an interesting problem that is not addressed by our design yet. This is a twofold problem as is explained below: 
\begin{enumerate}
\item Selecting friends that results in maximum availability.
\item In spite of limited resources available to portable devices, obtaining mirrors is competitive. Therefore, a user may need to propose incentives to potential candidates in order to motivate them to accept to be her mirrors. 

Providing the minimal incentives while increasing the chances of acceptance is analogous to the problem of bidding in online auctions.
\end{enumerate}

While our design addresses secure mirror synchronization amongst mirrors, when mirror synchronization should occur is another interesting challenge that was not part of our design and should be addressed in future works. Defining different incentives to motivate users to provide  rendezvous and relay services, and analyzing their effectiveness are research subjects that are worth investigation. A practical way of searching this peer to peer structure, while complying with all the user privacy and access control policies, is a very useful and yet hard to implement feature and is left out of our design for future works. 

Finally, in the near future, we intend to make MyZone available for large scale deployment. We plan to analyze its performance and the effectiveness of its security measures under different scenarios, both theoretically and experimentally.

\small
\bibliography{refs}
\bibliographystyle{plain}

\pagebreak

\end{document}